\documentclass[aps,prb,twocolumn,groupedaddress,showpacs]{revtex4-1}
\usepackage{graphicx}
\usepackage{amsmath}
\usepackage{relsize}
\usepackage{hyperref}
\usepackage{url}
\usepackage{multirow}
\usepackage{color,colortbl}
\usepackage{hhline}
\usepackage{float}
\usepackage[section]{placeins}
\usepackage[normalem]{ulem}
\restylefloat{table}

\begin{document}

\title{Thermodynamic parameters of single- or multi-band superconductors derived from self-field critical currents}

\author{E. F. Talantsev$^{1,*}$, W. P. Crump$^{1}$ and J. L. Tallon$^{1,2,*}$}

\affiliation{$^1$Robinson Research Institute, Victoria University of Wellington,
P.O. Box 33436, Lower Hutt 5046, New Zealand.}

\affiliation{$^2$MacDiarmid Institute for Advanced Materials and Nanotechnology,
P.O. Box 33436, Lower Hutt 5046, New Zealand.}

\affiliation{$^*$Correspondence and requests for materials should be addressed to E.F.T. (email:
Evgeny.Talantsev@vuw.ac.nz) or to J.L.T. (email: Jeff.Tallon@vuw.ac.nz)}

\date{\today}

\begin{abstract}
Key questions for any superconductor include: what is its maximum dissipation-free electrical current (its `critical current') and can this be used to extract fundamental thermodynamic parameters? Present models focus on depinning of magnetic vortices and implicate materials engineering to maximise pinning performance. But recently we showed that the {\it self-field} critical current for thin films is a universal property, independent of microstructure, controlled only by the penetration depth. Here we generalise this observation to include thin films, wires or nanowires of single- or multi-band $s$-wave and $d$-wave superconductors. Using extended BCS equations we consider dissipation-free {\it self-field} transport currents as London-Meissner currents, avoiding the concept of pinning altogether. We find quite generally, for type I or type II superconductors, the current is limited by the relevant critical field divided by the penetration depth. Our fits to 64 available data sets, from zinc nanowires to compressed sulphur hydride with critical temperatures of 0.65 to 203 K, respectively, are excellent. Extracted London penetration depths, superconducting energy gaps and specific heat jumps agree well with reported bulk values. For multiband or multiphase samples we accurately recover individual band contributions and phase fractions.
\end{abstract}

\pacs{74.25.Sv, 74.25.F-, 74.25.Bt, 84.71.Mn}

\maketitle

\section*{1. Introduction}

The occurrence of a critical current density, $J_c$, was initially proposed to be due to a depairing mechanism - the breaking of Cooper pairs by the Doppler shift arising from the velocity of the superfluid \cite{Volovik}.  However, the associated depairing critical current density, $J_{c,d}$, is so high for type II superconductors that it has never been convincingly demonstrated in experiments. There are still expectations, that in the case of nanofilaments, which are too small to contain a vortex, these high current densities may potentially be registered. %\cite{Bozovic}
But in larger samples (which until very recently encompasses about the entire inventory of data in the literature) this discrepancy seems to insist on different mechanisms of current limitation, chiefly those involved in Abrikosov flux pinning.  Even so, calculations of the energy gain due to Abrikosov vortex pinning \cite{Larbalestier} also lead to remarkably high $J_c$ values that were not observed in experiment.  Other vortex depinning mechanisms beyond simple Lorenz force depinning have been proposed, including thermally-activated depinning \cite{Gupta} together with other non-vortex mechanisms such as thermally-activated phase slip \cite{McCumber} and quantum phase slip \cite{Giordano1}. These have been applied variously for three dimensional (3D), quasi-two-dimensional (2D) and quasi-one-dimensional (1D) topologies.  Nevertheless, none of these mechanisms explain, still less provide a universal quantitative treatment for, the observed self-field critical current density, $J_c(\textrm{sf})$, in a wide variety of superconductors across these different topologies and over the full temperature range from near absolute zero up to $T_c$.

This may well be understandable, at least in terms of traditional vortex models where the pinning microstructure is inescapably variable. However, in the present work we show that, despite conventional expectations, this universal goal can be achieved using a London-Meissner model in which transport becomes dissipative when, for type I superconductors, the surface current density reaches the depairing magnitude, $B_c/(\mu_0\lambda)$ or, for type II superconductors, $B_{c1}/(\mu_0\lambda)$. Here $B_c$ is the thermodynamic critical field, $B_{c1}$ the lower critical field and $\lambda$ is the London penetration depth. We show that (i) the traditional vortex model for self-field $J_c$ is not quantitatively sustainable in many, possibly all, type II superconductors (see Section 2) and (ii) our non-vortex London-Meissner paradigm for self-field $J_c$ is quantitatively verified in every superconductor we have analysed (see all remaining sections). The self-field transport current is just the Meissner current which screens the self-field. For large samples this current is confined to within just a few $\lambda$ of the surface but for small samples the current profile extends over the full cross-section. The critical screening current corresponds to reaching a fundamental surface current-density limit (critical field divided by $\lambda$) and only for large samples does this correspond to a Silsbee-like surface-critical-field limit as we previously proposed \cite{Talantsev}.

For either superconductor type the model allows the self-field critical current density to be expressed purely in terms of $\lambda$ and, for type II superconductors, without any reference to vortex depinning. We use an extended Bardeen-Cooper-Schrieffer (BCS) model to fit a very wide range of data sets (64 in total) for $J_c(\textrm{sf})$. These fits allow key thermodynamic parameters (including the superconducting energy gap and the jump in electronic specific heat) to be determined in a manner which is simple and direct, in stark contrast to the more conventional methods for measuring these quantities. The test of the model is whether the fit parameters, especially the ground-state penetration depth, $\lambda_0$, correspond accurately to independently measured values. This test is astonishingly well met over all superconductor types and for $\lambda_0$ values extending over more than an order of magnitude.

\section*{2. The vortex problem}
\begin{figure}
\centering
\includegraphics[width=70mm]{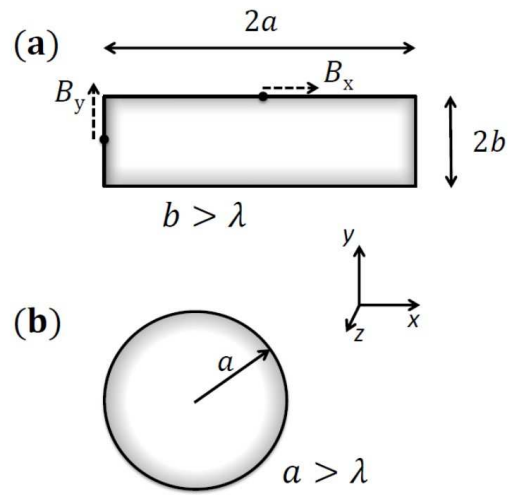}
\caption{\small
The cross-sectional geometry of the two conductors considered here: (a) films and (b) cylindrical wires, where in both cases current flows in the $z$-direction. The surface fields discussed the text, $B_x \equiv B_x(y=b)$ and $B_y \equiv B_y(x=-a)$, are illustrated. For films, the width in the $x$-direction is 2$a$ and the thickness in the $y$-direction is 2$b$, while for round wires the radius is $a$. For the case illustrated $a,b > \lambda$ and the transport current is just the Meissner shielding current flowing near the surface which screens the self-field. This is shown schematically by the shading. Only when $b<\lambda$ for films, or $a<\lambda$ for round wires, does the current flow across the entire cross-section and when $a,b \ll \lambda$ the current distribution is uniform. }
\label{geometry}
\end{figure}

The crucial deficiency in all presently available vortex models that aim to describe the self-field regime is the lack of quantitative fidelity. To demonstrate this we can consider the experimental data of Plourde {\it et al.} \cite{Plourde1}, where critical currents in amorphous MoGe thin films were reported. We use the geometry and axes shown in Fig.~\ref{geometry} with film width in the $x$-direction, film thickness in the $y$-direction and transport current in the $z$-direction.  Based on the most recent estimated values for the London penetration depth in MoGe films, $\lambda$(4.2 K) = 400 nm \cite{Latimer} and of the coherence length, $\xi_0 = 5$ nm \cite{Plourde1,MoGe}, we can calculate the London depairing critical current density, $J_d$, using:
\begin{equation}
J_d(T) = \frac{\phi_0 }{2\sqrt{2}\pi \mu_0 \xi(T) \lambda^2(T)}  . \label{depair}
\end{equation}
\noindent We obtain $J_d$(4.2 K) = 23.2 MA/cm$^2$ which is more than an order of magnitude larger than the measured $J_c$ values for MoGe \cite{Plourde1} (see Table I).

We can also calculate absolute values of the magnetic field produced by the transport current in these MoGe films using Ampere's law. The surface field, $B_x$, at the center of the film parallel to the large flat surfaces can be calculated from Ampere's equation to be:
\begin{equation}
B_x(y=\pm b) = \mp \mu_0 b J_c  , \label{Ampere}
\end{equation}
\noindent where $\mu_0$ is the permeability of free space. With reference to Fig.~\ref{geometry} the film width is 2$a$ and the film thickness is 2$b$. The calculated values for $B_x(y=b)$ for all MoGe films studied by Plourde {\it et al.} \cite{Plourde1} are presented in Table I. At the same time the magnetic field, $B_y(x=\pm a)$, at the film edges perpendicular to the flat surfaces can be calculated using Eq. (4) of Brojeny and Clem \cite{Brojeny}:
\begin{equation}
B_y(x=\pm a) = \mp \left(\mu_0 b J_c/\pi\right) \left[ \ln\left( \frac{2a}{b}\right) + 1 \right] . \label{Clem}
\end{equation}
\noindent Calculated values of $B_y(x=\pm a)$ for all these MoGe films are listed in Table I.

The thermodynamic critical field, $B_c$, can be calculated using
\begin{equation}
B_c(T) = \frac{\phi_0 }{2\sqrt{2}\pi \xi(T) \lambda(T)}  ,  \label{Thermo}
\end{equation}
\noindent while the lower critical field is
\begin{equation}
B_{c1}(T) = \frac{\phi_0 }{4 \pi \lambda(T)^2} \left(\ln\kappa +0.5\right)  .  \label{Lower}
\end{equation}

\begin{table*}
%\tiny
\centering
\begin{tabular}{|l||c|c|c|c|c|}
\hline
Parameter & Sample 1 & Sample 2 & Sample 3 & Sample 4 & Sample 5  \\ \hline \hline
 width, 2$a$ ($\mu$m) & 10 & 20 & 25 & 30 & 40  \\ \hline
 critical current, $I_c$ (mA) & 35 & 52.5 & 56.6 & 62.4 & 81.6  \\ \hline
 $J_c(\textrm{sf})$ (MA/cm$^2$) & 1.75 & 1.31 & 1.13 & 1.04 & 1.02  \\ \hline
 $B_x(y=\pm b)$ (mT) & 2.20 & 1.65 & 1.42 & 1.31 & 1.28  \\
 (surface field) &  &  &  &  & \\ \hline
 $B_y(x=\pm a)$ (mT) & 3.92 & 3.31 & 2.95 & 2.79 & 2.85  \\
 (edge field) &  &  &  &  & \\ \hline
 $\lambda_0$ (nm) & 332-339 & 366-373 & 384-392 & 395-403 & 397-405  \\
 (from $J_c$ using Eq.~\ref{type2}) &  &  &  &  & \\ \hline

\end{tabular}
\caption{Parameters for MoGe thin films (thickness 2$b$ = 200 nm) reported or calculated from Plourde {\it et al.} \cite{Plourde1}. Field values are all less than $B_{c} = 117$ mT and $B_{c1} = 5.03$ mT. The calculated range of values of $\lambda_0$ have the lower limit based on $\kappa = 80$ \cite{Plourde2} and the upper limit based on $\kappa = 108$ \cite{Plourde1}.  }
\label{MoGe}
\end{table*}

It can be seen from Table I that neither $B_x(y=\pm b)$ nor $B_y(x=\pm a)$ exceed the calculated $B_c$(4.2 K) = 117 mT or even $B_{c1}$(4.2 K) = 5.03 mT for these MoGe samples. Based on these calculations (which simply use Ampere's law) we can conclude that the magnetic field produced by the transport critical current in MoGe films is too low for vortices to exist in any area of the films at the conditions when the critical current is reached. By the ``critical current" we understand the conventional electric-field criterion of 1.0 $\mu$V/cm that was used also by Plourde {\it et al.} \cite{Plourde1}. The situation becomes worse still for thinner samples and becomes extreme in the case of few-monolayer superconductors, such as are increasingly being studied \cite{Zhang}. For each of the wide and disparate range of superconductors discussed below we calculate and list in the summary Tables (Appendix A) the values of $B_x$ and $B_y$ along with values of $B_c$ and, for type II superconductors, $B_{c1}$. In many cases the crucial edge field, $B_y$, is far less than $B_{c1}$ (both columns highlighted in gray) while in many other cases it is far more, suggesting there simply is no relationship between the two at $J_c$. Yet in every case, without exception, the surface current density is always equal to the critical field divided by $\lambda$. We particularly note the results listed in Table VII, Appendix A, for nanoscale samples where in every case $B_y \ll B_{c1}$. This point was actually made by Clem and coworkers \cite{Clem} in relation to a specific example of 22.5 nm thick NbN films (which we analyse in Fig.~\ref{sfits} and Table IV) where they ``conclude that the self-field and pinning effects are unimportant at all temperatures". Because our estimates are based on the simplicity of Ampere’s law this problem cannot be solved by assuming vortex pinning, because vortices do not yet exist at such low fields.

Our attempts to understand the physics of this central vortex problem led us to a model \cite{Talantsev} that describes the self-field critical current densities in all type-II thin-film superconductors (with thickness, 2$b < \lambda(0)$) by the universal equation given below as Eq.~\ref{type2}. In the following we show that this emerges from a universal critical surface current density, $J_s$, of the same magnitude and which is wholly analogous to the behavior seen in type I superconductors where vortices necessarily are absent but $B_{c1}$ is simply replaced by $B_c$. To place this in the context of the present data for MoGe we deduced the London penetration depth, $\lambda_0$, from the critical current for MoGe films using Eq.~\ref{type2}, below. In each case the range of values of $\lambda_0$ is listed in Table I with the lower limit based on the Ginsburg-Landau parameter $\kappa = 80$ \cite{Plourde2} and the upper limit based on $\kappa = 108$ \cite{Plourde1}. The deduced values are all very similar to the most recent estimated value for MoGe thin films, $\lambda(4.2$) K = 400 nm \cite{Latimer}. In every case of the large number of superconductors investigated below we find the same universal quantitative agreement between measured values of $\lambda_0$ and those deduced from self-field critical current measurements. This universal deficiency of the vortex model, on the one hand, together with the universal success of our transport current model, on the other, is the overriding motivation for the present work.

As discussed previously \cite{Talantsev}, vortex pinning starts to play a role in type II superconductors when an external magnetic field is applied to the superconductor and the flux front starts to propagate from the superconductor edges towards the center. Clearly this comes into play when the combined field at the edges (including the self-field and external field) exceed $B_{c1}$ sufficiently in order for vortices to nucleate, overcome surface barriers and migrate inwards under the Lorentz force across the bulk pinning field. Such considerations are beyond the scope of the present work.

\section*{3. Scope}

Because some of the present ideas have been introduced previously \cite{Talantsev} we briefly distinguish the scope of the present work from our earlier work. In that study we considered films of thickness comparable to $\lambda$ and found that, for all superconductors investigated, the global $J_c(\textrm{sf})$ (across the total cross-section) is given by the critical field divided by $\lambda$. This led us to infer a universal Silsbee-type criterion \cite{Silsbee} namely that dissipation sets in when the surface field reaches the critical field - applicable even for type II superconductors where the relevant critical field is $B_{c1}$. Here we consider samples of all sizes including $b\ll\lambda$ and $b\gg\lambda$ which enables us to find that, contrary to Silsbee, the critical threshold occurs in the surface current density, $J_s$, (and by Ampere's law a critical surface field-gradient) - not the surface field, $B_s$. For type I superconductors this critical $J_s$ is just the depairing current density, $B_c/(\mu_0\lambda)$. But for type II superconductors the critical $J_s$ is $B_{c1}/(\mu_0\lambda)$, a new an unexpected result. As a further new development we calculate $\lambda(T)$ from an extended BCS expression for the superconducting gap, $\Delta(T)$ so that by fitting the $T$-dependent $J_c(\textrm{sf},T)$ data sets we can extract $\lambda_0$, $T_c$, $\Delta_0$ and $\Delta C/C$ as free fitting parameters, where $\Delta C/C$ is the relative jump in electronic specific heat at $T_c$. We also extend to cylindrical symmetry; we investigate many more superconductors of various kinds; and we extend to multi-band and multi-phase superconductors. Some of the literature data sets are the same as used previously, but the analysis is different and it is valuable to have an essentially exhaustive analysis for all available superconductors in a single publication.

Fig.~\ref{geometry} shows the geometry of the two conductor types we consider here. Fig.~\ref{geometry}(a) shows a rectangular film of width, 2$a$, extending in the $x$-direction and thickness 2$b$, extending in the $y$-direction with total current, $I$, flowing in the $z$-direction. In most cases considered below $a \gg b$. Fig.~\ref{geometry}(b) shows a cylindrical wire of radius, $a$, with current, $I$, flowing in the $z$-direction. In the examples shown $a,b > \lambda$ so the transport current is the Meissner shielding current (shaded) flowing near the surface which screens the self-field.

\section*{4. Model}

As noted, the foundations of our approach were established recently \cite{Talantsev}. We analyzed $J_c(\textrm{sf})$ for thin films of half thickness, $b \approx \lambda$, in terms of transport London currents for which dissipation commences when the global current density reaches, for type I superconductors:
\begin{equation}
J_c(T,\textrm{sf}) = \frac{B_{c}(T)}{\mu_0\lambda(T)} = \frac{\phi_0 \kappa(T)}{2\sqrt{2}\pi \mu_0 \lambda^3(T)}  , \label{type1}
\end{equation}
\noindent while, for type II:
\begin{equation}
J_c(T,\textrm{sf}) = \frac{B_{c1}(T)}{\mu_0\lambda(T)} = \frac{\phi_0}{4\pi \mu_0 \lambda^3(T)} \left(\ln\kappa + 0.5\right) , \label{type2}
\end{equation}
\noindent where $\mu_0$ is the permeability of free space, $\xi$ is the superconducting coherence length and $\kappa = \lambda/\xi$ is the Ginsburg-Landau parameter which is only weakly $T$-dependent and is effectively constant under the logarithm in Eq.~\ref{type2}. Thus, leaving aside the variation of $\kappa$ with temperature this gives in both cases $J_c(\textrm{sf}) \propto \lambda^{-3} \equiv \rho_s^{3/2}$, where $\rho_s$ is the superfluid density. This allows absolute values of $\lambda(T)$ to be extracted from the $T$-dependence of $J_c(\textrm{sf})$ along with values of $\Delta_0$ from low-$T$ fits to $\lambda(T)$.

For the case when $b \approx \lambda$, these equations were shown to be well satisfied for a wide range of superconductors including metals, oxides, cuprates, heavy Fermions, ferro-pnictides, borocarbides, fullerenes and MgB$_2$. However, when $b > \lambda$ we showed that $J_c(\textrm{sf})$ adopts a thickness dependence which is well approximated by a multiplicative term $(\lambda/b)\tanh(b/\lambda)$ in Eqs.~\ref{type1} and ~\ref{type2}, as was confirmed by analysis of films of YBa$_2$Cu$_3$O$_y$ of varying thickness \cite{Talantsev}. This factor was introduced as a phenomenological approximation for the crossover function from small to large thickness, but in fact it proves to be exact as we see later. This factor takes the limits of $1$ for $b < \lambda$ and $(\lambda/b)$ for $b \gg \lambda$. Thus for $b \gg \lambda$ we have $J_c(\textrm{sf}) \propto \lambda^{-2} \equiv \rho_s$, and so with increasing size a crossover from $J_c(\textrm{sf}) \propto \lambda^{-3}$ to $J_c(\textrm{sf}) \propto \lambda^{-2}$ should be observed. We confirm this below in section 3.4. More generally, for a rectangular conductor of finite dimensions relative to $\lambda$ we may replace this size-dependent factor by  $\left[(\lambda/a)\tanh(a/\lambda) +(\lambda/b)\tanh(b/\lambda)\right]$, where $2a$ is the film width and, as before, 2$b$ is the film thickness. In summary, these equations for type-I and type-II superconductors, along with this thickness/width correction factor, allow us to use the same approach for quantifying $J_c(\textrm{sf})$ for all superconductors of rectangular cross-section without having to distinguish between different size or aspect ratios: thin-films, nanowires or bulk conductors.

Here we present a further generalization of the model based on BCS theory and the so-called $\alpha$-model \cite{Padamsee} for strong-coupling to show that absolute values of the basic thermodynamic parameters $\Delta(T)$, $\lambda(T)$, $\left(\Delta C/C\right)_{T=T_c}$ and $T_c$ can be extracted by fitting the self-field critical current density for single- and multi-band superconductors with various topologies, including cylindrical symmetry. We illustrate for a number of type I and type II superconductors of both $s$-wave and $d$-wave symmetry ranging from Zn nanowires with $T_c$ = 0.65 K to highly compressed H$_2$S with $T_c$ = 203 K, as recently reported \cite{H2S}. Moreover, for type I superconductors we find that irrespective of whether $b < \lambda$ or $b > \lambda$ the onset of dissipation under self-field transport occurs when the surface current density reaches the depairing current density, whereas it is only for $a,b \gg \lambda$ that this coincides with the surface field reaching $B_c$. For $b < \lambda$ the surface field never reaches $B_c$ before the depairing current limit is met. Similar behavior is inferred below for type II superconductors.

The BCS expression for the penetration depth in a flat-band weak-coupling $s$-wave superconductor is:
\begin{equation}
\left(\frac{\lambda(0)}{\lambda(T)}\right)^{2} = 1 - \frac{1}{2 k_B T} \int_{0}^{\infty} \! \cosh^{-2}\left( \frac{\sqrt{\varepsilon^2 + \Delta^2(T)}}{2 k_B T} \right) \,\mathrm{d}\varepsilon \label{BCSs}
\end{equation}
\noindent while for a $d$-wave superconductor with a 2D cylindrical Fermi surface it is:
\begin{align}
\left(\frac{\lambda(0)}{\lambda(T)}\right)^{2} &= 1 - \frac{1}{2 \pi k_B T} \int_{0}^{2\pi} \! \cos^2(\theta)  \nonumber \\
 &\times  \int_{0}^{\infty} \! \cosh^{-2}\left( \frac{\sqrt{\varepsilon^2 + \Delta^2(T,\theta)}}{2 k_B T} \right) \,\mathrm{d}\varepsilon \,\mathrm{d}\theta \label{BCSd}
\end{align}
\noindent  where $k_B$ is Boltzmann's constant and for $d$-wave symmetry $\Delta(T,\theta) = \Delta(T)\times \cos(2\theta)$ where $\theta$ is the angle around the Fermi surface subtended at ($\pi$,$\pi$) in the Brillouin zone. An analytical expression for the superconducting gap $\Delta(T)$ that allows for strong coupling is given by Gross \cite{Gross}:
\begin{equation}
\Delta(T) = \Delta(0) \, \tanh \left[\frac{\pi k_B T_c}{\Delta(0)} \sqrt{\eta \left(\frac{\Delta C}{C} \right) \! \left( \frac{T_c}{T} - 1 \right) } \, \right] \label{Delta}
\end{equation}
\noindent where $\Delta C/C$ is the relative jump in electronic specific heat at $T_c$. For $s$-wave symmetry $\eta = 2/3$ \cite{Gross} while, for $d$-wave symmetry using the Padamsee $\alpha$-model \cite{Padamsee}, we find that $\eta$ is well approximated by the ratio 7/5. By combining these equations we may fit the observed $J_c(T,\textrm{sf})$ data using $\lambda(0)$, $\Delta(0)$, $\Delta C/C$ and $T_c$ as free fitting parameters. We use non-linear curve fitting in the ORIGIN plot package for single-band $s$-wave fits, and for $d$-wave and multi-band fits we use the MATLAB software package (ver. R2012a, 32-bit (win32)). Our software is available for use on-line \cite{Wayne}.

\section*{5. Results}
\subsection*{3.1 $s$-wave superconductors}
\begin{figure}
\centering
\includegraphics[width=80mm]{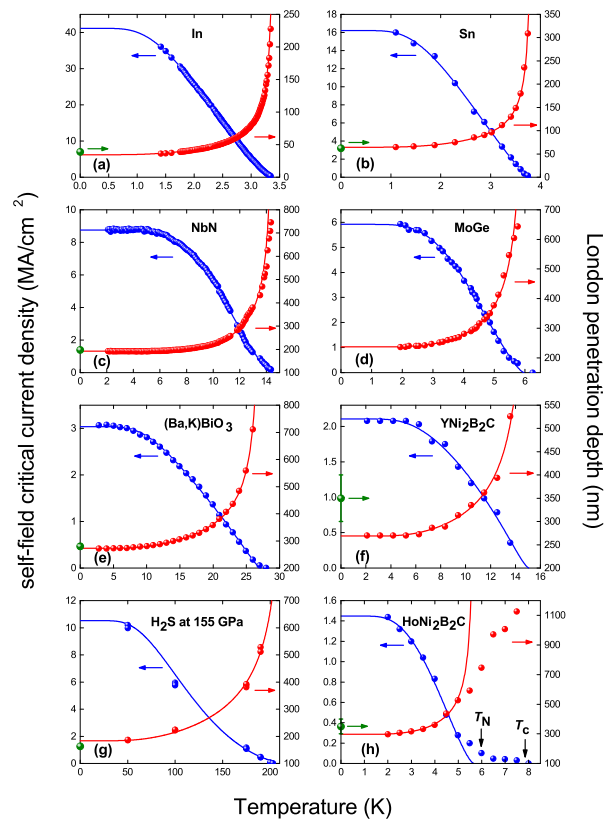}
\caption{\small
Experimental self-field $J_c(T)$ data for $s$-wave superconductors as annotated (left axis, blue) together with values of $\lambda$ (right axis, red) derived as described in the text. The solid curves are the BCS fits using Eqs.~\ref{type1} or~\ref{type2}, and~\ref{BCSs} and~\ref{Delta}. Note the variable offset of the $\lambda(T)$ axis. The single green data points at $T=0$ are, where available, reported ground-state values of $\lambda_0$ from our previous paper \cite{Talantsev}. Fit parameters: $\lambda(0)$, $\Delta(0)$, $\Delta C/C$ and $T_c$ are summarised in the Appendix Table IV. References for the raw $J_c(\textrm{sf})$ data sets are: (a) In \cite{Indium}, (b) Sn \cite{Sn}, (c) NbN \cite{NbN}, (d) MoGe \cite{MoGe}, (e) (Ba,K)BiO$_3$ \cite{BKBO}, (f) YNi$_2$B$_2$C \cite{YNi2B2C}, (g) H$_2$S at 155 GPa \cite{H2S} and (h) HoNi$_2$B$_2$C \cite{YNi2B2C}.  }
\label{sfits}
\end{figure}

We require strong-linked transport $J_c(\textrm{sf})$ data extending over a wide range of temperature, especially to low $T$ to enable accurate inference of $\Delta_0$. This limits the available data sets to relatively few. We show in Figure~\ref{sfits} most of what is available in the literature: six $s$-wave superconductors, of both type I and type II kind, including In \cite{Indium}, Sn \cite{Sn}, NbN \cite{NbN}, MoGe \cite{MoGe}, (Ba,K)BiO$_3$ \cite{BKBO}, YNi$_2$B$_2$C \cite{YNi2B2C}, H$_2$S at 155 GPa \cite{H2S} and HoNi$_2$B$_2$C \cite{YNi2B2C} where the stated references are for the raw $J_c(\textrm{sf})$ data sets. The associated fits using Eqs.~\ref{type1} or ~\ref{type2}, ~\ref{BCSs} and ~\ref{Delta} are also shown and in all cases the fits are excellent. $J_c(\textrm{sf})$ (blue) is shown on the left axis and $\lambda(T)$ (red) on the right. Free fitting values $\lambda(0)$, $\Delta(0)$, $\Delta C/C$ and $T_c$ are summarised in the Appendix Table IV, and independently reported ground-state values of $\lambda_0$, where available, are shown by the single dark green data points on the $T=0$ axis.

\begin{figure}
\centering
\includegraphics[width=80mm]{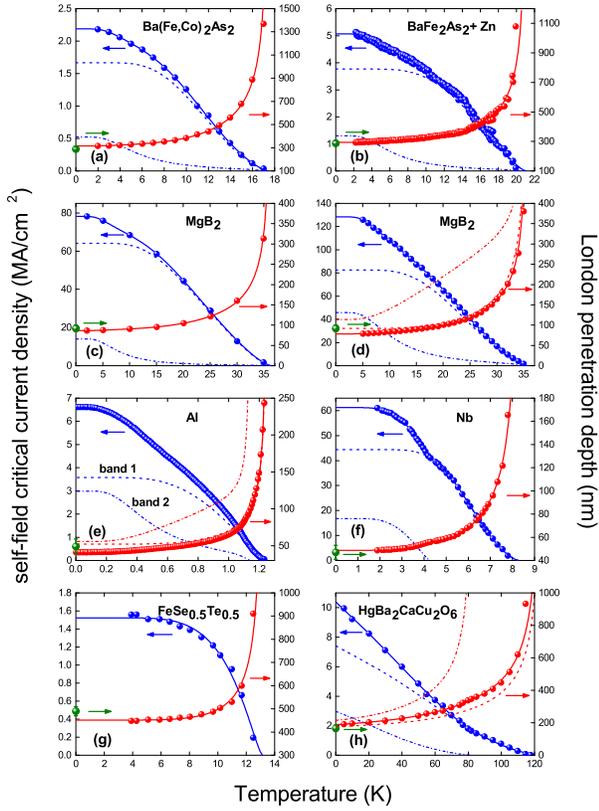}
\caption{\small
(a)-(h) $J_c(T,\textrm{sf})$ data for {\it two-band} superconductors (left axis, blue) with derived values of $\lambda$ (right axis, red).  Green data points at $T=0$ are reported ground-state values of $\lambda_0$ \cite{Talantsev}. Solid curves are fits using Eqs.~\ref{BCSs} and ~\ref{Delta} for each band where the partial $J_c$ values are combined linearly following Eqn.~\ref{twoband}. Dashed curves (dash/dot) show the partial contribution due to the first (second) band. Panels (a) to (d) are for strongly-coupled bands while (e) to (h) are for effectively decoupled bands. References for the raw $J_c(\textrm{sf})$ data sets are: (a) Ba(Fe,Co)$_2$As$_2$ \cite{BaFeCoAs}, (b) Ba(Fe,Zn)$_2$As$_2$ \cite{BaFeAsZn}, (c) MgB$_2$ \cite{MgB2b}, (d) MgB$_2$ \cite{MgB2a}, (e) Al \cite{Al2b}, (f) Nb \cite{MoGe} and (g) FeSe$_{1/2}$Te$_{1/2}$ \cite{FeSeTe}. Panel (h) shows data for a {\it two-phase} film of nominal composition HgBa$_2$CaCu$_2$O$_6$ (Hg1212) with intergrowths of Hg1201. Individual contributions are shown for 1212 (dashed) and 1201 (dash-dot).}
\label{twobfits}
\end{figure}

Immediately evident is the exponentially flat $T$-dependence at low-$T$ which is characteristic of the superfluid density of $s$-wave superconductors \cite{Talantsev}. On the whole there is excellent agreement between the deduced and reported values of $\lambda_0$. However it must be recognised that this parameter has not in every case been accurately determined and in particular its derivation from measured $B_{c1}$ values is fraught with problems \cite{Talantsev}. In addition to the data shown we carried out fits for five examples of MoGe and five of NbN and the representative plots shown for each of these in Figure~\ref{sfits} are typical of the wider data sets. The fit-parameters returned are all consistent with each other (see Appendix Table IV). We also fitted five examples of Al \cite{Al} using the single-band model and each of these returned values of $\lambda_0 \approx 50$ nm, close to the reported value. However, we found other $J_c(\textrm{sf})$ data that extended to much lower temperature and this allowed a two-band fit as shown later in Figure~\ref{twobfits}.

In Figure~\ref{sfits} the strong-coupling superconductors like NbN are evident from the fact that the flat $J_c$ region extends to higher temperatures followed by a more rapid fall on approaching $T_c$ where pair-breaking arising from the strong coupling imposes a lower-than-projected $T_c$ value. These also generally exhibit larger values of $(\Delta C/C)_{T=T_c}$. Panel (g) in the figure is for highly-compressed H$_2$S which was recently reported as having a record high $T_c$ of 203 K \cite{H2S}. The $J_c$ data is based on magnetisation measurements and was analysed in detail elsewhere \cite{Talantsev2}. For this system all parameter fit-values are surprisingly similar to the cuprates. The independently reported value for $\lambda(0)$ (= 163 nm, green data point) is from the authors' reported value of $B_{c1} = 30$ mT \cite{H2S,Talantsev2}. In their paper $\lambda(0)$ was quoted as 125 nm but this was in error \cite{Talantsev2}.  The last entry (h) is for HoNi$_2$B$_2$C and it should be compared with YNi$_2$B$_2$C shown above it. This reveals a two-step evolution of superconductivity with a weak onset at $T_c \approx 8$ K below which the superfluid density and $J_c$ rise slowly until the Ne\'{e}l temperature, $T_N \approx 6$ K, where long-range commensurate antiferromagnetic order is established (see arrows in the figure). Comparable two-step behavior in the development of the superconducting order parameter has also been reported for this compound \cite{Hill}.

\begin{figure}
\centering
\includegraphics[width=80mm]{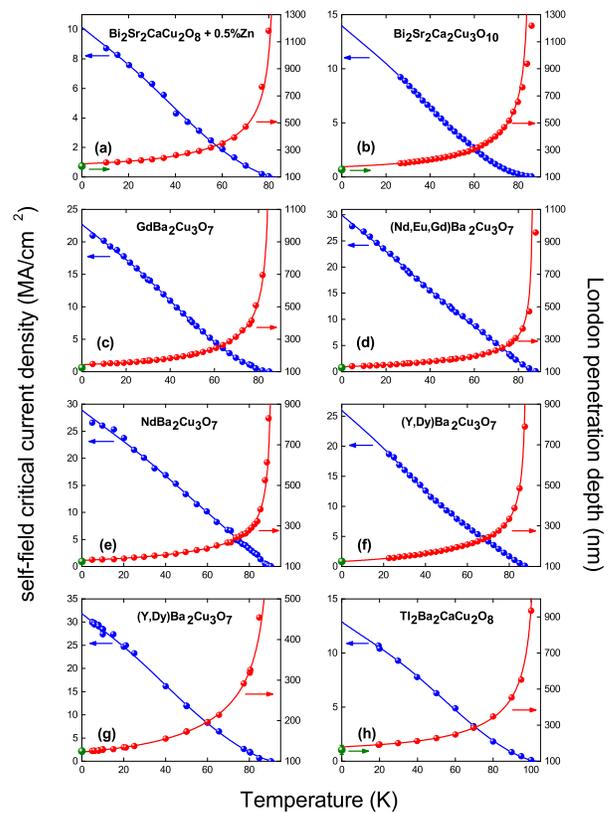}
\caption{\small
Experimental self-field $J_c(T)$ data for $d$-wave superconductors as annotated (left axis, blue) together with values of $\lambda$ (right axis, red) calculated as described in the text. The solid curves are the fits using Eqs.~\ref{BCSd} and ~\ref{Delta}. Note the offset of the $\lambda(T)$ axis by 100 nm. The single green data points at $T=0$ are reported ground-state values of $\lambda_0$ from our previous paper \cite{Talantsev}. Fit parameters: $\lambda(0)$, $\Delta(0)$, $\Delta C/C$ and $T_c$ are summarised in Appendix Table V.  References for the raw $J_c(\textrm{sf})$ data sets are: (a) Bi$_2$Sr$_2$CaCu$_2$O$_8$ \cite{Bi2212}, (b) Bi$_2$Sr$_2$Ca$_2$Cu$_3$O$_{10}$ \cite{Bi2223}, (c) GdBa$_2$Cu$_3$O$_y$ \cite{NdEuGd123}, (d) (Nd,Eu,Gd)Ba$_2$Cu$_3$O$_y$ \cite{NdEuGd123}, (e) NdBa$_2$Cu$_3$O$_y$ \cite{Nd123}, (f) (Y,Dy)Ba$_2$Cu$_3$O$_y$ \cite{YDy123}, (g) (Y,Dy)Ba$_2$Cu$_3$O$_y$ \cite{Talantsev}, and (h) Tl$_2$Ba$_2$CaCu$_2$O$_8$ \cite{Tl2212}.  }
\label{dfits}
\end{figure}

\subsection*{3.2 $d$-wave cuprates}

Similarly, we show in Figure~\ref{dfits} $J_c$ data for eight cuprates along with their associated fits using Eqs.~\ref{type2}, ~\ref{BCSd} and ~\ref{Delta}. Again $J_c(T,\textrm{sf})$ (blue) is shown on the left axis and $\lambda(T)$ (red) on the right, and the fitting values for $\lambda(0)$, $\Delta(0)$, $\Delta C/C$ and $T_c$ are summarised in Appendix Table V. Note the offset of 100 nm on the right-hand $\lambda(T)$ axis. Again, in all cases the fits are excellent now showing the linear-in-$T$ low-temperature behavior which is distinctive of $d$-wave symmetry and which sharply contrasts the behavior seen in Figure~\ref{sfits} for the $s$-wave examples. In all cases the ground-state values, $\lambda(0)$, are in excellent agreement with literature values (green data points at $T=0$ and also listed in Appendix Table V). For all the RBa$_2$Cu$_3$O$_y$ samples we used the reported value $\lambda_0 = 125$ nm as discussed in our previous paper \cite{Talantsev}. The one data set available for HgBa$_2$CaCu$_2$O$_6$ is considered later as an example of two-phase intergrowths showing the efficacy of $J_c$ measurements in identifying such intergrowths.

\subsection*{3.3 Multiple bands}

Turning now to two-band superconductivity, various groups have calculated the specific heat \cite{Bouquet} and superfluid density \cite{Carrington} where the two distinct band contributions (to the specific heat or superfluid density) are combined linearly in proportion to the partial Sommerfeld constants, $\gamma_i$, of each band. We may call this the $\gamma$-model (as distinct from the $\alpha$-model mentioned above). For two-band contributions to $J_c(\textrm{sf})$ the appropriate combination is perhaps less clear. In the macroscopic case where $J_c(\textrm{sf})$ scales as $\lambda^{-2}$ it is clear that
\begin{equation}
J_c(\textrm{tot}) = \gamma J_c(1) + (1-\gamma)J_c(2) \label{twoband}
\end{equation}
\noindent just like the linear combination of partial superfluid densities. Here each partial $J_c$ is calculated from each partial superfluid density and $\gamma = \gamma_1/(\gamma_1 + \gamma_2)$ where $\gamma_i$ are the partial Sommerfeld constants. However, when the conductor dimension is smaller than $\lambda$ and $J_c(\textrm{sf})$ scales as $\lambda^{-3}$ we do not have a precedent.

We argue here that if we were merely to add superfluid densities and determine $J_c$ from the total combined superfluid density this would result in one channel reaching its critical state at $J_c$ while the other channel remains sub-critical. We take the view that in the critical state both channels will be at their respective critical thresholds. The onset of dissipation in one channel spills current density onto the other sub-critical channel until they are both critical. Thus it is the partial $J_c$ values that must be linearly combined not the partial superfluid densities, and Eq.~\ref{twoband} applies in the general case irrespective of whether $J_c(i)$ scales as $\lambda(i)^{-2}$ or $\lambda(i)^{-3}$ for band $i$. For strongly-coupled bands, for each band we use Eq.~\ref{Delta} in combination with Eqs.~\ref{BCSs} or ~\ref{BCSd} using for each band the same $T_c$ value \cite{Suhl} and the same total $\lambda_0$ value. We also fix $\Delta C/C$ for the lowest-energy-gap band at the BCS value of 1.43 for $s$-wave symmetry. Thus the free fitting parameters are then $\gamma$, $T_c$, $\lambda_0$, $\Delta_0(1)$, $\Delta_0(2)$ and $\Delta C(1)/C(1)$. For effectively decoupled bands we allow independent band values for $T_c$ and for $\lambda_0$ and free-fit all parameters: $T_c(1)$, $T_c(2)$, $\lambda_0(1)$, $\lambda_0(2)$, $\Delta_0(1)$, $\Delta_0(2)$, $\Delta C(1)/C(1)$ and $\Delta C(2)/C(2)$.

Figure~\ref{twobfits} (a) to (d) shows the data and their associated overall fits (solid curves) for Ba(Fe,Co)$_2$As$_2$, Ba(Fe,Zn)$_2$As$_2$ and MgB$_2$ for which two examples are shown. These are examples of strongly-coupled bands. For effectively decoupled bands (e) to (h) then show Al, Nb, FeSe$_{1/2}$Te$_{1/2}$ and, lastly, HgBa$_2$CCu$_2$O$_6$ which will be discussed later. The dashed curves (dash/dot) show the partial contribution due to the first (second) band. Note again the variable offset of the $\lambda(T)$ axis; and again, literature values for $\lambda_0$ listed in our previous work \cite{Talantsev} are shown by the single green data points at $T=0$. They exhibit exceedingly good agreement with $\lambda(T)$ calculated from the overall $J_c(\textrm{sf},T)$ fits. All fit parameters are listed in Appendix Table VI. To avoid clutter we only show the partial contributions to $\lambda(T)$ from each band in panels (d) and (e). Even though FeSe$_{1/2}$Te$_{1/2}$ is a two-band superconductor the data does not extend to low enough temperature to complete a two-band analysis. The fit here is a single-band fit, but the contribution from the second band may possibly already be evident in the small rise in $J_c$ at the lowest temperature.

Multi-band behavior is most evident in the continuing rise in $J_c(\textrm{sf})$ at lower temperatures in contrast to the broad, exponentially flat behavior seen in single-band $s$-wave superconductors. The magnitudes of the two gaps, as reported in Appendix Table VI, are largely consistent those reported from measurements using tunneling or ARPES. We do not find evidence in the literature of two superconducting gaps in the case of aluminium but the transport behavior is indeed governed by two bands as shown long ago \cite{Ashcroft}. The two examples for MgB$_2$ are included to show the overall consistency among distinct data sets from different groups. Some minor differences are evident but the overall magnitude of the two gaps, as evidenced by the temperature scales of the partial $J_c$ contributions, are entirely consistent. A second data set for MgB$_2$ in ref.\cite{MgB2a} gives an almost identical fit to that shown in Fig.~\ref{twobfits}(d).

\subsection*{3.4 Different topologies}

We now extend the application of our equations to cylindrical symmetry and compare three different practical topologies: (i) 3D round wires, (ii) quasi-2D thin films with $b \leq \lambda$ (as already discussed above), and (iii) quasi-1D nanowires of cross-sectional dimension $2a,2b \leq \lambda$ in order to demonstrate the different scaling behaviors of these topologies.

The case of cylindrical symmetry was discussed by London and London \cite{London1}. The local current-density distribution at radial position $r$ in a cylindrical wire of radius $a$ carrying a total current $I$ along its axis in the $z$-direction is
\begin{equation}
J(r) = \frac{i I}{2\pi a \lambda} \frac{J_0(ir/\lambda)}{J_1(ia/\lambda)} \equiv \frac{I}{2\pi a \lambda} \frac{I_0(r/\lambda)}{I_1(a/\lambda)}  , \label{London}
\end{equation}

\noindent where $i \equiv \sqrt{-1}$, $J_0(x)$ and $J_1(x)$ are zeroth- and first-order Bessel functions of the first kind and  $I_0(x)$ and $I_1(x)$ are zeroth- and first-order {\it modified} Bessel functions of the first kind. When $a \gg \lambda$ this is usually replaced by the standard exponential decay in $\textbf{J}$ at the surface. But when $a \leq \lambda$ the Bessel function solution results in a rather (but not precisely) uniform distribution in $\textbf{J}$. The field distribution is given by the London equation $B = -\mu_0 \lambda^2 \left(\partial J/\partial r\right)$ so that, quite generally:
\begin{equation}
B(r) = \frac{\mu_0 a}{2} J_{\textrm{av}} \frac{I_1(r/\lambda)}{I_1(a/\lambda)}  , \label{Field}
\end{equation}

\noindent where $J_{\textrm{av}} = I/(\pi a^2)$ is the global-average current density. The field at the surface is $B_s = (1/2)\mu_0 a J_{\textrm{av}}$ while the local current density at the surface is $J_s = J_{av}(a/2\lambda)\times I_0(a/\lambda)/I_1(a/\lambda)$. In our fits we always use the exact Bessel function solutions but for those interested in convenient approximations we have $J_s \approx J_{av}(a/2\lambda)\times \left[\tanh(a/2\lambda)\right]^{-1}$ which is a good approximation across the entire range of $a/\lambda$ and is asymptotically exact in both the small and large limits. For a type I superconductor dissipation is expected to set in either when $B_s \rightarrow B_c$ or when $J_s \rightarrow J_{c,d}^L = \phi_0\kappa/(2\sqrt{2}\pi \mu_0 \lambda^3)$, the London depairing current density. It is easy to see that for $a \gg \lambda$ these dissipation onset criteria are identical. But for $a \leq \lambda$ the current-density limit is reached before the field limit. So in general we focus only on the {\it local} current limit $J_s \rightarrow J_{c,d}^L$ for which the critical value, $J_c$, of $J_{\textrm{av}}$ is
\begin{align}
J_c(\textrm{sf}) &=  J_{c,d}^L \left(\frac{2\lambda}{a}\right) \times \frac{I_1(a/\lambda)}{I_0(a/\lambda)} \label{Jcwiregen} \\
 &\approx J_{c,d}^L \left(\frac{2\lambda}{a}\right) \tanh\left(\frac{a}{2\lambda}\right) . \label{Jcwireprox}
\end{align}

\begin{figure}
\centerline{\includegraphics*[width=60mm]{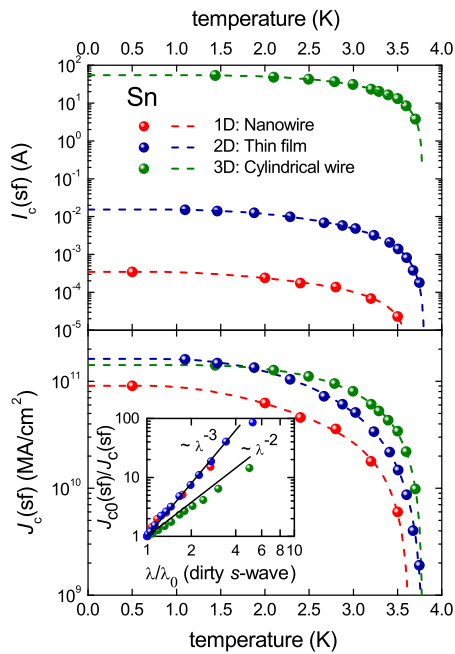}} \caption{\small
(a) Self-field critical current and (b) critical current density for Sn in three conductor topologies (i) annular film on round quartz rod \cite{Hagedorn}, (ii) thin film \cite{Sn} and (iii) nanowire with radius $a$ = 35 nm \cite{Tian}. The dashed curves are BCS-like fits using Eqs. ~\ref{BCSs} and ~\ref{Delta}. The inset in (b) shows a log-log plot of normalized $J_c^{-1}$ versus normalized $\lambda$ with $T$ as the implicit variable. The solid lines show the ideal cubic and quadratic relationships as annotated. }
\label{Snfits}
\end{figure}

\noindent We note that $\left(2\lambda/a\right) \tanh\left(a/2\lambda\right)$ effectively takes the value 1 for all $a < \lambda$ and not just for $a \ll \lambda$, thus leaving $J_c \approx J_{c,d}^L$ pretty much over the entire sub-penetration-depth length scale. Thus $J_c \propto \lambda^{-3}$ for $a < \lambda$ and, effectively, $J_c \propto \lambda^{-2}$ for $a \gg \lambda$, as follows:
\begin{align}
J_c &=  \frac{\phi_0 \kappa}{2\sqrt{2} \pi \mu_0 \lambda^3} & a < \lambda  \label{Jcwirelt} \\
 &=  \frac{\phi_0 \kappa}{\sqrt{2} \pi \mu_0 a \lambda^2} & a \gg \lambda  \label{Jcwiregt}
\end{align}
\noindent We show below (Figure~\ref{Snfits}) that these inverse cube and square limits are indeed realised in the experimental data. (We also note that there may be a case for replacing the London depairing current density, $J_{c,d}^L$, in Eq.~\ref{Jcwiregen} by the slightly smaller Ginzburg-Landau depairing current density, $J_{c,d}^{GL}$, given by
\begin{equation}
J_{c,d}^{GL} = \frac{\phi_0 \kappa}{3\sqrt{3}\pi \mu_0 \lambda^3} . \label{JcGL}
\end{equation}
\noindent with exactly the same scaling behavior, but we do not consider this further.)

In the case of type II superconductors we originally posited that the self-field $J_c$ criterion occurs when the surface field reaches $B_{c1}$ \cite{Talantsev}. There we were considering thin films of half thickness $b \approx \lambda$. As a consequence of this thinness we suggested that the logarithmically diverging attraction of vortices of opposite sign nucleated on opposite faces of the film ensures that surface barriers and pinning are surmounted such that the vortices are indeed nucleated as soon as $B_{c1}$ is reached, they migrate inwards under the combination of Lorenz force and mutual attraction, then annihilate at the centre. Hence the inevitable onset of dissipation is reached. Moreover, we concluded that this interaction lowers the energy of formation of vortices thus reducing $B_{c1}$ by the factor $\tanh(b/\lambda)$. However, this picture is not sustainable when $b \leq \xi$ nor for macroscopic samples where surface barriers and pinning must begin to play a role.

Instead, we can only conclude that the self-field $J_c$ is, as for type I, determined by a fundamental surface current density limit, not a fundamental surface field limit. And this is found to be the case equally for macroscopic samples and nanoscopic samples where the dimension is so small that vortices are non-existent. As already noted this fundamental surface current limit for type II superconductors is $J_s = B_{c1}/(\mu_0\lambda)$. For cylindrical symmetry this leads to:
\begin{align}
J_c(\textrm{sf}) &= \frac{\phi_0}{2\pi \mu_0 \lambda^2 a}(\ln \kappa + 0.5) \times \frac{I_1(a/\lambda)}{I_0(a/\lambda)} \nonumber \\
&= J_c^* \left(\frac{2\lambda}{a}\right) \times \frac{I_1(a/\lambda)}{I_0(a/\lambda)}  \label{cylT2gen}\\
&\approx J_c^* \left(\frac{2\lambda}{a}\right) \times \tanh\left(\frac{a}{2\lambda}\right) , \label{cylT2genprox}
\end{align}
\noindent where $J_c^*$ is given precisely by Eq.~\ref{type2} and we recall this $J_c(\textrm{sf})$ is the global $J_c$ over the full cross-section. This equation, Eq.~\ref{cylT2gen}, is quite general and applies for large or small $a$. Indeed experimentally we find it holds over a huge range of values of $a/\lambda$ and we conclude on observational grounds that type II superconductors are current-limited, not field-limited, just as we found rigorously for type I superconductors. Again, as in type I superconductors the current limit coincides with the field limit for $a \gg \lambda$ but not when $a \leq \lambda$. Because Eq.~\ref{cylT2gen} is identical in form to Eq.~\ref{Jcwiregen} one might even say that Eq.~\ref{type2} (i.e. $J_c^*$) acts like an effective depairing $J_c$ for type II superconductors. Though the origins of this are not yet clear this observation is a major conclusion of this work.  As in the type I case discussed above, this relation, Eq.~\ref{cylT2gen}, also has a $\lambda^{-2}$ or $\lambda^{-3}$ limiting behavior depending on whether $a \gg \lambda$ or $a < \lambda$. In the former case the limiting behavior is asymptotic to the field limit while in the latter case the surface field falls well short of the bulk $B_{c1}$. And in either case $J_c$ falls well below the London depairing limit, $J_{c,d}^L$, which is higher by a factor of the order of $\sqrt{2}\kappa/(\ln\kappa + 0.5)$.

So to conclude, for cylindrical conductors, of any size, dissipation sets in when the surface current density, $J_s$ reaches $B_{c}/(\mu_0\lambda)$ for type I superconductors (the London depairing limit) or $B_{c1}/(\mu_0\lambda)$ for type II superconductors (for reasons yet to be fully established) and the relevant equations are ~\ref{Jcwiregen} and ~\ref{cylT2gen}. As noted, these will be shown to be satisfied over a huge range of $a/\lambda$.

Before analysing data sets for round wires in the light of these equations we return briefly to the implications of the above thinking for thin films. In the critical state the local current density only varies in the $y$-direction so the solution to the London equation leads to
\begin{equation}
J(y) = J_s \frac{\cosh(y/\lambda)}{\cosh(b/\lambda)}  , \label{Londfilm}
\end{equation}
\noindent where $J_s$ is the surface current density, top or bottom. Replacing $J_s$ by $B_{c1}/(\mu_0\lambda)$ for the type II critical current criterion and integrating Eq.~\ref{Londfilm} from $-b$ to $+b$ to get the total critical current over the cross-section we find
\begin{equation}
J_c(\textrm{sf}) = (\lambda/b) \tanh(b/\lambda) \frac{B_{c1}}{\mu_0\lambda}  . \label{Jcfilm}
\end{equation}
\noindent And a similar relation holds with $B_{c1}$ replaced by $B_c$ for type I. This now shows that our previously phenomenological $(\lambda/b) \tanh(b/\lambda)$ factor \cite{Talantsev} is indeed exact under our current-limit criterion.

%Having clarified this point we turn finally to the case of a cylindrical annular film of outer radius $a$ and inner radius $d \gg \lambda$. Applying Ampere's law at the surface ($r = a$) tells us that $B_s \times 2\pi a = \mu_0 I = \mu_0 J_{\textrm{av}}\times 2\pi a (a-d)$. For a type I superconductor the critical value of $J_{\textrm{av}}$ is reached when $B_s = B_c$. Thus
%\begin{align}
%J_c &=  \frac{\phi_0 \kappa}{2\sqrt{2} \pi \mu_0 (a-d) \lambda^2} ; & (a-d) \geq \lambda , \ d \gg \lambda . \label{Jcannul}
%\end{align}
%\noindent When $(a-d) \gg \lambda$ Meissner currents flow only in a thin layer of the outer surface of the cylinder and decay exponentially to zero towards the inner surface of the cylinder. It is easy to show that when $B_s \rightarrow B_c$ then at the same time $J_s \rightarrow J^L_{c,d}$. Moreover, this result is identical to Eq.~\ref{Jcwiregt} except that here the current density is referred to the annular cross-section whereas in Eq.~\ref{Jcwiregt} the current density is referred to the total cylindrical cross-section.

We now use Eqs.~\ref{Delta} and ~\ref{BCSs} or ~\ref{BCSd} in combination with these $\lambda$-dependent expressions for $J_c$ to extract the thermodynamic parameters for some representative type I superconductors by fitting the $T$-dependence of $J_c(\textrm{sf})$ for these different topologies. Again, the fitting parameters are $T_c$, $\Delta_0$, $\left(\Delta C/C \right)_{T=T_c}$ and $\lambda_0$.

We focus on the case of Sn for which we have three examples. Hagedorn \cite{Hagedorn} presents $I_c(T,\textrm{sf})$ data for a 170 nm thick annular film of Sn on a round fused quartz rod of diameter 719 $\mu$m.  The relevant equation is Eq.~\ref{Jcwiregen}. Hunt \cite{Sn} gives $I_c(T,\textrm{sf})$ data for a thin film with $a$ = 950 nm and $b$ = 25 nm, while Tian {\it et al.} \cite{Tian} give $I_c(T,\textrm{sf})$ data for a nanowire with radius, $a$ = 35 nm. These data embrace 3D, quasi-2D and quasi-1D conductors for which the relevant equations are ~\ref{Jcwiregen}, ~\ref{type1} and ~\ref{Jcwirelt}, respectively (though in the last case we use the exact Eq.~\ref{Jcwiregen} to fit data for round cross-section). In particular we expect $J_c(T,\textrm{sf})$ in the first case to scale as $\lambda^{-2}$ and in the latter two cases as $\lambda^{-3}$. We will see this is borne out. The raw $I_c(T,\textrm{sf})$ data is plotted for these three systems in Fig.~\ref{Snfits}(a) and for $J_c(T,\textrm{sf})$ in Fig.~\ref{Snfits}(b). While the self-field critical currents for these cases range over seven orders of magnitude, the current densities collapse to essentially the same scale. By using $\kappa$ = 0.23 \cite{Poole12} the fits as described above were performed and are plotted as the dashed curves in Fig.~\ref{Snfits}.

The inset in (b) shows a log-log plot of $J_c^{-1}$ versus $\lambda$ with $T$ as the implicit variable and the solid lines show the ideal cubic and quadratic relationships as annotated. Evidently, the film and nanowire closely follow the cubic relationship indicated by Eqs.~\ref{type1} and ~\ref{Jcwirelt} while the annular film more or less follows the quadratic relationship indicated by Eq.~\ref{Jcwiregt} - as  predicted. Fit values of $T_c$, $\Delta_0$, $\left(\Delta C/C \right)_{T=T_c}$ and $\lambda_0$ are listed in Appendix Table VI for the three geometries. These fit values are generally in good agreement with each other and with the reported values listed underneath.

Four more data sets for Sn films are available from Song and Rochlin \cite{Song} and these span a range of film thicknesses from $b$ = 49 nm to 298 nm. These reveal what the authors interpreted from the different $T$-dependences as a crossover from bulk transport to Josephson transport. Unfortunately absolute $J_c(\textrm{sf})$ values are only available for two of the samples and these can only be found in Song's thesis \cite{Song}. These are for $b$ = 49 nm and $b$ = 190 nm. Our fits, again, are excellent and they show a very clear crossover from $J_c(\textrm{sf}) \propto \lambda^{-3}$ for the thinnest ($b\approx\lambda$) to $J_c(\textrm{sf}) \propto \lambda^{-2}$ for the thicker film ($b\gg\lambda$), precisely as expected from Eqs.~\ref{Jcwirelt} and ~\ref{Jcwiregt}. In our view this is the origin of the change in $T$-dependence and probably has nothing to do with granular Josephson transport.

\begin{figure}
\centering
\includegraphics[width=80mm]{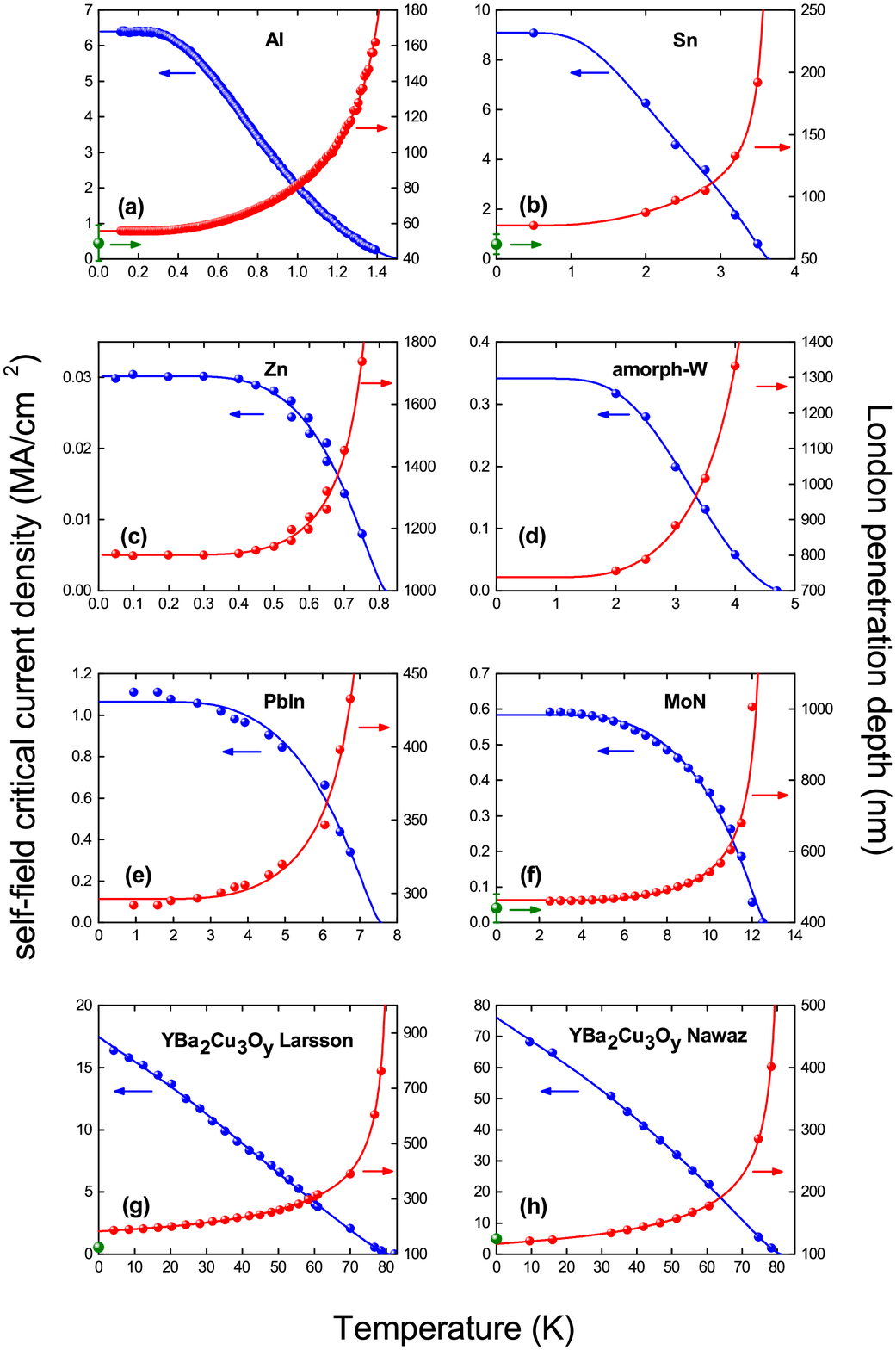}
\caption{\small
Experimental self-field $J_c(T)$ data for a variety of $s$- and $d$-wave superconducting nanowires as annotated (left axis, blue) together with values of $\lambda$ (right axis, red) calculated as described in the text. The solid curves are the fits using Eqs.~\ref{BCSs} and ~\ref{Delta}. Fit parameters: $\lambda(0)$, $\Delta(0)$, $\Delta C/C$ and $T_c$ are summarised in the Appendix Table VII. References for the raw $J_c(\textrm{sf})$ data sets are: (a) Al \cite{Alnano}, (b) Sn \cite{Tian}, (c) Zn \cite{Znnano}, (d) amorphous-W \cite{AmorphW}, (e) PbIn \cite{PbInnano}, (f) $\delta$-MoN \cite{MoNnano}, (g) YBa$_2$Cu$_3$O$_7$ \cite{Larsson} and (h) YBa$_2$Cu$_3$O$_7$ \cite{Nawaz}. Where available independently-measured ground-state values of $\lambda_0$ are shown by the green data points at $T=0$ K \cite{Talantsev}. }
\label{nanowires}
\end{figure}

\definecolor{Gray}{gray}{0.9}
\newcolumntype{g}{>{\columncolor{Gray}}c}

\begin{table*}
%\tiny
\centering
\begin{tabular}{|l||c|c|c|c|c|c|c|g|g|c|}
\hline
Material & $2a $ & $2b$ & I$_c$ (2 K) & J$_c$(2 K) & $\lambda$ (2 K) & $\lambda$(0) (nm) & $B_x$ & $B_y$  & $B_{c1}$  & B$_c$  \\
& $\mu$m & $\mu$m & A & MA/cm$^2$ & (nm) & independently  & (mT) & (mT) & (mT) & (mT)  \\
&  &  &  &  &  & measured  &  &  &  &    \\ \hline \hline
 In (cylindrical) \cite{indium3D} & 520 & - & 22.4 & 0.0105 & 38.5 & 40 \cite{Poole12} &  17.2 & - & - & 17.3 \\
 $\kappa$ = 0.11 \cite{Poole12} & & & & & &  &  &  &  & \\ \hline
  & 270 & - & 12.1 & 0.0211 & 37.8 &  & 17.9 & - & - & 17.9 \\ \hline
  & 170 & - & 8.15 & 0.036  & 36.5 &  & 19.2 & - & - & 19.2 \\ \hline
  & 170 & - & 7.88 & 0.0347 & 37.2 &  & 18.5 & - & - & 18.5 \\ \hline \hline
Material & $2a $ & $2b $ & I$_c$ (4.2 K) & J$_c$(4.2 K) & $\lambda$ (4.2 K) & $\lambda$(0)  (nm)  &  &  &  &  \\
 & $\mu$m & $\mu$m & A & MA/cm$^2$ & (nm) & independently  &  &  &  &    \\
 & &  &  &  &  & measured  &  &  &  &    \\ \hline \hline
Nb \cite{Huebener} & 82 & 1 & 4.2 & 5.12 & 50.9  & 47 \cite{Maxfield}, 59 \cite{Masuradze} & 32.2 & 52.3 & 31.8 & 89.9 \\
$\kappa$ = 1.0 \cite{Kim1} &  & &  &  &   & 41 $\pm$ 4 \cite{Felcher}  &  &  &  & \\ \hline
 & 76 & 1 & 3.5 & 4.61 & 53.7 & & 29.0 & 46.4 & 28.6 & 80.8 \\ \hline
 & 93 & 1 & 4.1 & 4.41 & 54.8 & & 27.7 & 46.1 & 27.4 & 77.6 \\ \hline
 & 62 & 1 & 2.4 & 3.87 & 58.6 & & 24.3 & 37.4 & 23.9 & 67.6 \\ \hline
 & 49 & 1 & 2.9 & 5.92 & 47.5 & & 37.2 & 54.4 & 36.5 & 103  \\ \hline
 & 83 & 1 & 4.2 & 5.06 & 51.2 & & 31.8 & 51.8 & 31.4 & 88.9 \\ \hline
 Nb$_3$Sn \cite{Cody}  & 101 & - & 44.9 & 0.56 & 56.7 & 65 \cite{Poole12}  & 178 & 178 & 184 & 1,590 \\
 (cylindrical) &  & &  &  &   &   &  &  &  & \\
 $\kappa$ = 22 \cite{Poole12} &  & &  &  &  & & &  & & \\ \hline
 (cylindrical) & 94 & - & 31.9 & 0.46 & 66 &  & 136 & 136 & 136 & 1,180 \\ \hline
 & 150 & 36 & 41.4 & 0.761 & 65.2 &  & 172 & 122 & 139 & 1,210 \\ \hline
 &  & & 43.4 (2.1 K) & 0.804 (2.1 K ) & 63.5 (2.1 K) &  & 182 & 129 & 147 & 1,270 \\ \hline

\end{tabular}
\caption{Critical currents and penetration depth of Nb, Nb$_3$Sn and In measured at 4.2, 2.1 and 2 K. $B_x$ and $B_y$, defined in Fig.~\ref{geometry}, are the surface field components at the middle of the flat surface and edge, respectively. $B_{c1}$ and $B_c$ are calculated from $\lambda_0$. The $B_y$ and $B_{c1}$ columns are highlighted in gray for ease of comparison.}
\label{NbNbSn}
\end{table*}

\subsection*{3.5 Nanowires}

We could not find any other data available with which to compare these three topologies and so we focus now on other reported nanowire systems. In Figure~\ref{nanowires} we assemble the data for eight different nanowire conductors: Al, Sn, Zn, amorphous-W, PbIn, $\delta$-MoN and two examples of YBa$_2$Cu$_3$O$_7$. The fits are done using Eqs.~\ref{Jcwiregen} or~\ref{cylT2gen}, depending on whether the system is type I or type II, respectively, together with Eqs.~\ref{BCSs} or ~\ref{BCSd} and ~\ref{Delta}. For nanowires with rectangular cross-section we use the thickness-width correction factor $\left[(\lambda/a)\tanh(a/\lambda) +(\lambda/b)\tanh(b/\lambda)\right]$. Again, the fit parameters, listed in the Appendix Table VII, are in good agreement with literature values and independently measured ground-state values of $\lambda_0$ are shown in Figure~\ref{nanowires} by the single green data points on the $T=0$ axis \cite{Talantsev}. For $\delta$-MoN the value $\lambda_0 = 440 \pm 40$ nm has been reported \cite{MoN}, again in excellent agreement with our value. On the other hand, reported values for PbIn: $\lambda_0 = 150$ nm \cite{Poole12} is off scale but this value should be regarded as tentative only, and if there are defects in the nanowires $J_c$ tends to be low and the inferred $\lambda_0$ is too high. There appears to be no data available for amorphous-W. In the case of YBa$_2$Cu$_3$O$_7$ the first example due to Larsson {\it et al}., \cite{Larsson} exhibits a rather low $J_c$ (and along with this, a high value of $\lambda_0$) but this system is notoriously dependent upon full oxygenation and on the absence of impurity scattering which rapidly suppresses $\rho_s$ and $J_c$. The second example due to Nawaz {\it et al}. \cite{Nawaz} exhibits $J_c(\textrm{sf})$ at low-$T$ as high as 70 MA/cm$^2$. Bearing in mind the additional factor of two in Eq.~\ref{cylT2gen} for nanowires compared with Eq.~\ref{type2} for films, this is precisely consistent with the best quality films displaying $J_c(\textrm{sf}) \approx$ 30 - 35 MA/cm$^2$ - see Figure~\ref{dfits}(g). (This factor of two also arises from the thickness-width correction factor for rectangular cross-section superconductors).

We note again the columns in Table VII for the surface and edge fields, $B_x$ and $B_y$, and the values of $B_{c1}$ and $B_c$. It is evident that the edge field is far less than $B_{c1}$ by up to a factor of 40 (both columns highlighted in gray for comparison) and as a consequence vortices can play no role in the onset of dissipation in any of these samples yet, again, in all cases the surface current density is $B_{c1}/(\mu_0 \lambda)$ as is found for all other samples investigated in Tables I through VII, including the very largest samples as described in the next section.

\subsection*{3.6 Bulk wires}

For bulk, round indium wire (type I) we used Eq.~\ref{Jcwireprox} to fit the data. For the few examples of bulk, round, type II wires we use Eq.~\ref{cylT2gen} (or its approximation Eq.~\ref{cylT2genprox}) combined with Eqs.~\ref{BCSs} and ~\ref{Delta} to fit the data which is reported at 4.2 K or 2.1 K, only. The fit parameters are summarised in Table II for Nb and Nb$_3$Sn, including also bulk rectangular samples as indicated by the dimensions in columns 2 and 3. Notably, the values of $\lambda_0$ returned for Nb with thickness 1 micron are the same as those obtained for Nb thin films of dimension 50 times smaller (see Fig.~\ref{twobfits}). Moreover, the round wire Nb$_3$Sn conductors, as much 100 times larger, also return values of $\lambda_0$ in full agreement with the independently measured value. This is a powerful validation of our model and in particular Eq.~\ref{cylT2gen}, namely that from samples with dimension as small as 5 nm (Al) to as large as 100 $\mu$m (Nb$_3$Sn), more than four orders of magnitude, our analysis of $J_c(T,\textrm{sf})$ faithfully yields values of the London penetration depth and its $T$-dependence.

\subsection*{3.7 Multiple phases, intergrowths}

Multiple phases, intergrowths and decoupled bands are treated in the same manner as described in section 3.3 for multiple bands but, where possible, with $T_c(1)$, $T_c(2)$, $\lambda_0(1)$ and $\lambda_0(2)$ as additional free fitting parameters. In particular, and to illustrate, many of the layered cuprate high-$T_c$ superconductors are members of homologous series with more or less the same $a$- and $b$-axes but having different numbers of CuO$_2$ planes per unit cell and hence different $c$-axes. They therefore often exhibit intergrowths of one homologue in a matrix of another. These phases have different intrinsic $T_c$ values and will therefore exhibit a more complex $T$-dependence of the superfluid density and, according to the above, a more complex $T$-dependence of $J_c(\textrm{sf})$.  An illustrative example is shown in Figure~\ref{twobfits}(h) where $J_c(T,\textrm{sf})$ data is displayed for a film of nominal composition HgBa$_2$CaCu$_2$O$_6$ (Hg1212) \cite{Krusin}. This has a significant fraction of intergrowths of HgBa$_2$CuO$_4$ (Hg1201) as is evidenced by the two-step behavior in $J_c(T)$.

%\begin{figure}
%\centering
%\includegraphics[width=60mm]{Figure7.eps}
%\caption{\small
%Self-field critical current density, $J_c(\textrm{sf})$, for a film of nominal composition HgBa$_2$CaCu$_2$O$_6$ (Hg1212) but possessing a fraction of intergrowths of Hg1201. The data is from Krusin-Elbaum \cite{Krusin} and is fitted as a linear combination of partial $J_c$ contributions from each phase (see text) using $J_c = f J_c(1212) + (1-f) J_c(1201)$. The $J_c$ contribution from each phase is each fitted using Eqs.~\ref{type2}, ~\ref{BCSd} and ~\ref{Delta}. As there are now nine fit parameters, including the fraction $f$, $\Delta C/C$ for 1201 is fixed at the value 1.4. Individual contributions are shown for 1212 (red) and 1201 (blue) as well as the overall fit to the measured $J_c$ (black).}
%\label{Hgfit}
%\end{figure}

We fit this data using, for each phase, the formulae for type II $d$-wave films, namely Eqs.~\ref{type2}, ~\ref{BCSd} and ~\ref{Delta}, and we enforce weak-coupling fits to reduce the free-parameter set as the available raw data is rather limiited. The contributions of each phase are added in the same way as the two-band linear combination summarised in Eq.~\ref{twoband} where $\gamma$ is now the fraction of the dominant phase and $(1-\gamma)$ the fraction of intergrowth phase. The difference here is that for strongly-coupled two-band behavior there is a single $T_c$ value whereas for two-phase behavior there are two distinct $T_c$ values, so $T_c(1)$ and $T_c(2)$ are now free fitting parameters. Additionally, there are two $\lambda(T)$ functions for each phase and it is not possible to separate $\gamma$, $\lambda_0(1)$ and $\lambda_0(2)$ by curve fitting. We thus use the linear Uemura relationship between $T_c$ and $\rho_s(0)$ \cite{Uemura} to set $\lambda_0(2) = \lambda_0(1) \sqrt{T_c(1)/T_c(2)}$, and proceed using $\gamma$ and $\lambda_0(1)$ as free fitting parameters (along with $T_c(1)$ and $T_c(2)$).

The data of Fig.~\ref{twobfits}(h) is analyzed in this way. The individual contributions are shown by the dashed (1212) and dash-dot (1201) curves giving an excellent overall fit to the reported $J_c(T,\textrm{sf})$ and we obtain the intergrowth phase fractions ($\gamma$ and $(1-\gamma)$) of 58\% Hg1212 and 42\% Hg1201. The original paper does not provide information to check whether these fractions are independently supported. However, the film preparation was modeled on the method of Tsui {\it et al}., \cite{Tsui} for which x-ray diffraction indeed revealed a significant fraction of Hg1201 present in the Hg1212 matrix. Moreover, early studies on apparently x-ray phase-pure Bi and Tl cuprates often showed from high-resolution transmission electron microscopy the presence of a high percentage of homologous intergrowths. Another candidate system for such studies is YBa$_2$Cu$_3$O$_y$ with intergrowth stacking faults of YBa$_2$Cu$_4$O$_8$, as is commonly observed.

We note that the deduced fraction $\gamma = 0.58 \pm 0.25$ has a large uncertainty with the present data set. But the fit is only presented as an illustration of two-phase analysis. To extract truly robust parameters one needs high-quality, closely-spaced data - significantly better than the present data set \cite{Krusin}. This is no criticism. Until now it has not been considered possible to extract fundamental quantities from $J_c$ data so groups have not worried too much about the fine detail thinking that this is just a matter of small variations in e.g. pinning profile. This said, we repeat our qualification noted earlier \cite{Talantsev} that under extremely strong pinning our London-Meissner model for {\it self-field} $J_c$ could possibly be eclipsed by more traditional vortex depinning, though we have not found any examples in our rather exhaustive analysis. Thus, what we present here is a simple new tool for structural/phase characterization of superconducting materials via $J_c$ measurements that does not involve microscopy or diffraction techniques.

\section*{6. Concluding remarks - vortices and pinning}

The above examples, combined with our previous work \cite{Talantsev}, show quantitatively that our proposed London-Meissner mechanism for self-field transport currents in type II superconductors brings into question the more traditional vortex displacement models. In all cases, for small or large sample dimensions with respect to $\lambda$, dissipation sets in when the surface Meissner shielding current density, $J_s$, reaches $B_{c1}/(\mu_0\lambda)$. For large samples this coincides with the surface field just reaching the magnitude of $B_{c1}$ but leaves no room for $B_s$ to exceed this value in order for vortices to nucleate, overcome surface barriers and migrate through the field of pinning microstructure. For small samples ($b\ll\lambda$) the surface and edge fields are actually far too small for vortices to even exist. Consider for example the analysis of 5 nm films of NbN presented in Fig.~\ref{sfits}(c) and in Table IV. At self-field critical current the edge field is $0.054\times B_{c1}$ and the surface field is $0.022\times B_{c1}$. Or consider the perhaps more interesting and compelling case of PbIn cylindrical nanowires \cite{PbInnano} presented in Fig.~\ref{nanowires}(e) and Table VII. Here there is no edge field, just a surface field calculated rigorously to be 0.09 mT, a factor of 45 below $B_{c1}$.

In other, much larger, samples the edge field substantially exceeds $B_{c1}$. There is therefore no correlation whatsoever between the edge field and $J_c$(sf), with values of the former varying over a huge range while, for the more than 80 samples we have investigated, the surface current density is always equal to the critical field divided by $\lambda$. This fact means that the vortex model is quantitatively unsustainable, while in every case the model we present here is validated.

For type II superconductors it is perhaps useful to derive the film thickness for which, at $J_c$(sf), the edge field reaches the magnitude of $B_{c1}$ . By combining Eq.~\ref{Clem} with Eq.~\ref{Jcfilm}, below, we can state quite generally (including anisotropic superconductors) that $B_y \leq B_{c1}$ when:
\begin{equation}
\left[ \ln\left( \frac{2a}{b}\right) + 1 \right] \tanh(b/\lambda_c) \leq \pi   , \label{criterion}
\end{equation}
\noindent where $\lambda_c$ is the $c$-axis penetration depth normal to the film. Thus, if for example $a = 100\times b$ then $B_y \ll B_{c1}$ if $b \leq 0.548\lambda_c$. Below this value the edge field drops as $\tanh(b/\lambda_c)$ i.e. essentially linearly in $(b/\lambda_c)$. Even if $b$ substantially exceeds the critical value given by Eq.~\ref{criterion} the field penetrating at the edges falls to the value of $B_{c1}$ in a very short distance. To see this we use equations (8) or (13) of Brojeny and Clem \cite{Brojeny} to obtain the $y$-component of the field on the long axis inside the film i.e. at the point $(x,0)$ where $x = a-\Delta x$:
\begin{equation}
B_y(x,0) = (1/2\pi) \tanh(b/\lambda_c) B_{c1}\times \ln\left( \frac{a+x}{a-x}\right)^2 . \label{BrojandClem}
\end{equation}
This field falls to the value of $B_{c1}$ within $\Delta x$ where
\begin{equation}
1 = (1/\pi) \tanh(b/\lambda_c)\times \ln\left( \frac{2a}{\Delta x}\right) . \label{BrojandClem}
\end{equation}
\noindent Thus when $b = \lambda_c$ then $\Delta x = 1.6$ \% of the film width; when $b = 2\lambda_c$ then $\Delta x = 3.8$ \% of the film width; and for only slightly thicker films $\Delta x$ saturates at 4.3 \% ($e^{-\pi}$) of the film width. As shown by Brandt and Indenbom \cite{Brandt} the field $B_y(x,0)$ then falls abruptly to zero and there is no further flux penetration beyond $\Delta x$. Thus, under self-field, for very thin or thick samples at $J_c(\textrm{sf})$ vortices of any sort appear to be irrelevant and yet the same surface current density limit occurs for our thinnest samples as for our largest macroscopic samples. If this is universal behavior then it would seem that vortices and vortex pinning are playing no role in self-field critical currents.

With the introduction of a perpendicular magnetic field the superfluid density will fall \cite{Sonier2} and along with it the magnitude of $J_c$ but this will be a small effect compared with the eventual vortex entry at the edges with the resultant change in current density profile as described by Brandt and Indenbom \cite{Brandt}. There will thus be a crossover from our model to more conventional models based on vortex entry and drift. An approach to describing this crossover will be discussed elsewhere.

Then at a more practical level, studies on pinning enhancements in self-field show no increase in $J_c(\textrm{sf})$. As discussed previously \cite{Talantsev} these studies include electron-beam-induced columnar arrays, neutron-irradiation studies and $^{16}$O ion-beam irradiation. In each case while there was significant increase in the in-field $J_c$ there was no increase in $J_c(\textrm{sf})$ \cite{Talantsev} (indeed there was a small decrease consistent with a small reduction in superfluid density). It would not be appropriate to reproduce these arguments in detail here, we simply close with an analysis of the so-called second-generation (2G) cuprate HTS conductors in which considerable effort has been devoted to enhancing pinning.

From a variety of different HTS 2G-wires, we analyse some representative samples that, for all except one, we have measured on an instrument described by Strickland and co-workers \cite{Y123ours}. Because in each sample $b \ll a$ we analysed the $J_c(\textrm{sf},T)$ data using Eq.(12) of our previous work \cite{Talantsev}:
\begin{equation}
J_c(\textrm{sf},T) = \frac{\phi_0(\ln \kappa + 0.5)}{4 \pi \mu_0\lambda_{ab}^{3}} \times (\lambda_c/b) \tanh(b/\lambda_c)   . \label{Jcfilm}
\end{equation}
\noindent and for simplicity we assume that $\lambda_c(T \approx 20 \textrm{K})$ = 1000 nm for all samples.

\begin{table*}
%\tiny
\centering
\begin{tabular}{|l||c|c|c|c|c|}
\hline
Parameter & (NEG)BCO \cite{NdEuGd123} & SuperPower tape \cite{Selv} & SuperPower tape \cite{Selv} & YBCO \cite{Talantsev} & STI tape \cite{Knibbe}  \\ \hline \hline
 2$a$ ($\mu$m) & 50 & 50 & 250 & 500 & 500  \\ \hline
 2$b$ (nm) & 50 & 1000 & 1000 & 850 & 4500  \\ \hline
 temperature (K) & 19 & 18 & 18 & 20 & 23  \\ \hline
 $I_c$ (A) & 0.61 & 11 & 56 & 105 & 240  \\ \hline
 $J_c(\textrm{sf})$ (MA/cm$^2$) & 24.3 & 21.5 & 22.4 & 24.7 & 10.6  \\ \hline
 derived $\lambda$ (nm) & 140 & 142 & 140 & 139 & 140  \\ \hline
 $B_x$ (mT) Eq.~\ref{Ampere} & 7.6 & 135 & 140 & 132 & 300  \\ \hline
 \rowcolor{Gray}
 $B_y$ (mT) Eq.~\ref{Clem} & 18.5 & 199 & 278 & 297 & 516  \\ \hline
 \rowcolor{Gray}
 $B_{c1}$ (mT) from $\lambda$ & 42.7 & 41.5 & 42.7 & 43.0 & 42.7 \\ \hline
 $B_{c}$ (mT) from $\lambda$ & 1130 & 1100 & 1130 & 1150 & 1130 \\ \hline

\end{tabular}
\caption{Parameters for YBCO 2G wires with calculated values for surface and edge fields. For calculations we assumed $\kappa = 95$ and $\lambda_c = 1000$ nm. The SuperPower and STI tapes are commercial products. The former has a high density of BaZrO$_3$ artificial pinning centres while STI tapes nominally have none. The YBCO is our in-house tape with a high density of Dy$_2$O$_3$ pinning centres. }
\label{MoGe}
\end{table*}

The first sample is a 50 nm thick $\times$ 50 $\mu$m wide (NEG)BCO film reported by Cai {\it et. al.} \cite{NdEuGd123}.  It has no inclusions of nanoparticles that typically are used as vortex pinning centres.  The deduced $\lambda(T=19 \textrm{K}) = 140$ nm (see Table III) is as expected. The calculated fields are such that $B_x < B_y < B_{c1}$ as expected for a thin and reasonably narrow film.

The second sample is a commercial SuperPower wire \cite{Selv} which we patterned to a bridge with the same width 2a = 50 m, but this wire has a 20 times thicker superconducting layer and, in addition, it contains a very high density of BaZrO3 nanoparticles. In spite of the associated enhanced pinning at high fields the self-field behavior is unaltered and we deduce $\lambda$ = 142 nm, which is (within the uncertainty of film thickness) essentially the same value as the pin-free (NEG)BCO film \cite{NdEuGd123}.

In the next sample the five-fold increase in bridge width 50 $\mu$m to 250 $\mu$m for the same SuperPower wire simply causes an increase in the $B_y$ field, as one can expect from Eq.~\ref{Clem}. All deduced parameters remain essentially unchanged despite the increase in edge field. In a vortex nucleation and migration model this increase should lower $J_c$(sf). It does not.

The next, 4th sample, is a YBCO film we prepared using the standard American Superconductor Corporation metal-organic deposition technique \cite{Rupich} and which has high density of Dy$_2$O$_3$ nanoparticles \cite{Xia}.  The deduced and calculated parameters for this sample are clearly identical to those found for the (NEG)BCO and SuperPower samples.

Finally we studied a commercial wire manufactured by Superconductor Technology Incorporated (STI) which is free of artificial pinning centers. This wire was studied by high-resolution transmission electron microscopy and “in-field” critical current measurements were made by other members of our group \cite{Knibbe}. It was confirmed that the wire does not contain any artificial pinning nanoparticles and, moreover, there is relatively low density of other types of defects that might serve as pinning centres. Despite this both the deduced $\lambda(T=23$ K) = 140 nm and $B_{c1}(T = 23$ K) = 42.7 mT are in excellent agreement with all other 2G-wire samples in Table III. Notably, $B_y$ reaches the unprecedented value of 0.5 T.

These results show that in spite of the completely different manufacturing processes, chemical compositions, sample thickness (factor of 90), sample width, types of pinning centers (or their absence) and widely variable self-fields, $B_x$ and $B_y$, the deduced fundamental parameters of these films ($\lambda$, $B_{c1}$ and $B_c$) are essentially the same. This again confirms that vortex nucleation and pinning is irrelevant to the magnitude of $J_c(\textrm{sf})$ and a more universal mechanism limiting $J_c(\textrm{sf})$ is operating here, having a fundamental rather than engineering nature.

\section*{7. Summary}

In summary, we have used generalized BCS expressions to calculate the temperature dependence of the superconducting energy gap in the more general case not necessarily confined to weak-coupling. From this the superfluid density is calculated. We use these expressions for non-linear fitting of self-field critical current density including type I and type II, $s$-wave and $d$-wave superconductors in thin film, nanowire and macroscopic 3D wire topologies - a total of 73 different data sets. We find that in all cases $J_c$ is governed by a universal surface-current density limitation, not a surface-field limitation, namely that $J_s \rightarrow B_c/(\mu_0\lambda)$ (type I) or $J_s \rightarrow B_{c1}/(\mu_0\lambda)$ (type II). The origin of the former is clear - this is the depairing current density. The precise origin of the latter is not - it is presented as an experimental observation which may lead to some interesting new physics.  In all cases we obtain excellent fits to the data and to the inferred temperature dependence of the penetration depth. The fitting parameters $T_c$, $\lambda_0$, $\Delta_0$ and $\Delta C/C_{T=T_c}$ are in good agreement with literature values and, in particular, inferred $\lambda_0$ values lie well within the range of independently reported values. The method is reliant on high-quality $J_c$ data extending from the lowest temperatures (to infer accurate $\Delta_0$ values) to very close to $T_c$ (to infer accurate $\Delta C/C$ values) on samples which are free of weak links. We extend these ideas to two-band superconductors and multi-phase superconductors and again find excellent parameter fits. The self-field critical current density is therefore a window into the key thermodynamic parameters in all superconductors and represents a very simple technique for extracting their values. In particular, our approach in deriving $\Delta_0$ is based on a measurement of the entire sample volume (when $b < \lambda_0$). Thus, surface artifacts are less likely to affect our analysis in comparison with other techniques such as scanning tunneling microscopy and angle resolved photoelectron spectroscopy, especially for highly reactive materials such as FeSe$_{1/2}$Te$_{1/2}$.”

JLT thanks the Marsden Fund of New Zealand and the MacDiarmid Institute for Advanced Materials and Nanotechnology for financial support.

\section*{APPENDIX A}
\subsection*{Tables of Parameter Values}

\begin{table*}[h]
\tiny
\centering
\begin{tabular}{|l||c|c|c|c|c|c|c|c|c|c|g|g|c|}
\hline
Material/ & $2a $ & $2b $ & $\kappa$ & $J_c(0)$ & $T_c$ & $\Delta_0$ & $\Delta C/C$ & $\lambda_0$ & 2$\Delta_0/k_B T_c$ & $B_x$ & $B_y$ & $B_{c1}$ & $B_c$ \\
geometry & nm & nm & & MA/cm$^2$ & (K) & (meV) &  & (nm) &  & (mT) & (mT) & (mT) & (mT) \\ \hline \hline
{\bf $s$-wave} & & & &  &  &  &  &  & & & & & \\ \hline

Al & 610 & 89 & 0.03 \cite{Poole12} & 4.44 \cite{Al} & 1.2 $\pm$ 0.01 & 0.211 $\pm$ 0.005 & 1.1 $\pm$ 0.07 & 49.3 $\pm$ 0.09 & 4.1 $\pm$ 0.1 & 2.5 & 2.1 &  & 2.9 \\
 & 680 & 98 & & 3.38 \cite{Al} & 1.18 $\pm$ 0.01 & 0.22 $\pm$ 0.01 & 1.6 $\pm$ 0.2 & 53.8 $\pm$ 0.2 & 4.3 $\pm$ 0.2 & 2.1 & 1.8 &  & 2.4 \\
 & 880 & 99 & & 4.06 \cite{Al} & 1.18 $\pm$ 0.01 & 0.22 $\pm$ 0.01 & 1.6 $\pm$ 0.2 & 49.2 $\pm$ 0.2 & 4.3 $\pm$ 0.2 & 2.5 & 2.4 &  & 2.9 \\
 & 500 & 34 & & 3.82 \cite{Al} & 1.25 $\pm$ 0.01 & 0.175 $\pm$ 0.002 & 1.5 $\pm$ 0.1 & 55.8 $\pm$ 0.08 & 3.25 $\pm$ 0.06 & 0.8 & 0.9 &  & 2.2 \\
 & 300 & 20 & & 3.68 \cite{Al} & 1.34 $\pm$ 0.01 & 0.186 $\pm$ 0.002 & 1.4 $\pm$ 0.1 & 59.3 $\pm$ 0.09 & 3.22 $\pm$ 0.06 & 0.5 & 0.5 &  & 2.0 \\
experiment & & & & &  & 0.179 \cite{Wolf} & 1.45 \cite{Poole4} & 50 $\pm$ 10 \cite{Prozorov} & & & & &  \\
  & & & & &  &  &  & 46-51 \cite{Doezema} 51.5 \cite{Biondi} & & & & & \\ \hline

In & 360 & 100 & 0.11 \cite{Poole12} & 41.1 \cite{Indium} & 3.378 $\pm$ 0.002 & 0.487 $\pm$ 0.003 & 1.181 $\pm$ 0.009 & 34.14 $\pm$ 0.06 & 3.35 $\pm$ 0.01 & 25.8 & 17.3 &  & 22.0  \\
 & 320 & 100 & & 20.9 \cite{Indium} & 3.489 $\pm$ 0.006 & 0.505 $\pm$ 0.006 & 1.21 $\pm$ 0.02 & 46.4 $\pm$ 0.1 & 3.36 $\pm$ 0.01 & 13.1 & 8.4 &  & 11.9 \\
experiment & & & & &  & 0.525 \cite{Poole7}, 0.541 \cite{Wolf} & 1.9 \cite{Bryant} & 40 \cite{Poole12} & & & & & \\ \hline

Sn nanowire & 70 & 70 & 0.23 \cite{Poole12} & 9.1 \cite{Tian} & 3.70 $\pm$ 0.04 & 0.54 $\pm$ 0.05 & 2.5 $\pm$ 0.6 & 77.0 $\pm$ 0.3 & 3.4 $\pm$ 0.3 & 2.0 & 2.0 &  & 9.0 \\
Sn film & 1,900 & 50 & & 16.2 \cite{Sn} & 3.78 $\pm$ 0.03 & 0.58 $\pm$ 0.03 & 2.0 $\pm$ 0.2 & 64.1 $\pm$ 0.8 & 3.5 $\pm$ 0.1 & 5.1 & 7.0 &  & 13.0  \\
Sn annular film & 718800 & 170 &  & 0.0136 \cite{Hagedorn} & 3.78 $\pm$ 0.01 & 0.75 $\pm$ 0.12 & 2.5 $\pm$ 0.3 & 41.8 $\pm$ 0.9 & 4.6 $\pm$ 0.7 & 30.7 & 30.7 &  & 30.7  \\
Sn film & 500 & 98 & & 30.4 \cite{Song} & 3.89 $\pm$ 0.03 & 0.57 $\pm$ 0.01 & 2.31 $\pm$ 0.26 & 51.7 $\pm$ 0.2 & 3.4 $\pm$ 0.1 & 18.7 & 14.4 &  & 20.0  \\
Sn film & 500 & 380 & & 18.3 \cite{Song} & 3.76 $\pm$ 0.01 & 0.58 $\pm$ 0.01 & 2.48 $\pm$ 0.09 & 46.5 $\pm$ 0.1 & 3.59 $\pm$ 0.06 & 43.7 & 17.9 &  & 24.8  \\
experiment & & & & &  & 0.593 \cite{Wolf} & 1.6 \cite{Bryant} & 56-68 \cite{Peabody} & & & & &   \\ \hline

NbN & 8,900 & 8 & 40 \cite{Poole12} & 7.9 \cite{Engel} & 14.37 $\pm$ 0.03 & 3.00 $\pm$ 0.03 & 2.83 $\pm$ 0.08 & 193.5 $\pm$ 0.1 & 4.83 $\pm$ 0.05 & 0.4 & 1.0 & 18.4 & 249 \\
 & 4,900 & 8 & & 8.65 \cite{Engel} & 13.62 $\pm$ 0.04 & 3.13 $\pm$ 0.04 & 2.67 $\pm$ 0.09 & 189.7 $\pm$ 0.1 & 5.33 $\pm$ 0.07 & 0.4 & 1.0 & 19.2 & 259  \\
 & 2,900 & 8 & & 8.76 \cite{Engel} & 14.50 $\pm$ 0.05 & 3.51 $\pm$ 0.08 & 1.85 $\pm$ 0.07 & 192.1 $\pm$ 0.1 & 5.6 $\pm$ 0.2 & 0.4 & 0.9 & 18.7 & 253  \\
 & 1,900 & 8 & & 8.02 \cite{Engel} & 13.79 $\pm$ 0.06 & 3.22 $\pm$ 0.08 & 2.0 $\pm$ 0.1 & 197.8 $\pm$ 0.1 & 5.4 $\pm$ 0.2 & 0.4 & 0.8 & 17.6 & 238  \\
 & 300 & 8 & & 14.3 \cite{Engel} & 13.85 $\pm$ 0.07 & 2.03 $\pm$ 0.03 & 1.7 $\pm$ 0.1 & 191.5 $\pm$ 0.3 & 3.46 $\pm$ 0.05 & 0.7 & 1.0 & 18.8 & 254  \\
 & 6,000 & 22.5 & & 7.47 \cite{Clem} & 11.81 $\pm$ 0.01 & 2.11 $\pm$ 0.08 & 1.98 $\pm$ 0.04 & 198.6 $\pm$ 1.3 & 4.15 $\pm$ 0.15 & 1.1 & 2.1 & 17.5 & 236  \\
experiment & & & & &  & 2.56 \cite{Kihlstrom} & 1.9 \cite{Geibel} $\pm$ 0.09 & 200 \cite{Poole12} 194 \cite{Komiyama} & & & & & \\ \hline

MoGe & 2,000 & 64 & 80 \cite{Plourde2} & 5.8 \cite{MoGe} & 5.90 $\pm$ 0.03 & 1.35 $\pm$ 0.12 & 1.86 $\pm$ 0.14 & 239 $\pm$ 2 & 5.3 $\pm$ 0.5 & 1.3 & 1.7 & 9.2 & 214 \\
 & 5,000 & 64 &  & 2.9 \cite{MoGe} & 6.28 $\pm$ 0.07 & 1.4 $\pm$ 0.3 & 2.1 $\pm$ 0.4 & 291 $\pm$ 7 & 5.2 $\pm$ 1.1 & 0.8 & 1.2 & 7.1 & 164 \\
 & 7,000 & 64 &  & 2.9 \cite{MoGe} & 6.30 $\pm$ 0.09 & 1.2 $\pm$ 0.1 & 1.9 $\pm$ 0.3 & 292 $\pm$ 3 & 5.7 $\pm$ 1.0 & 0.7 & 1.3 & 7.1 & 165 \\
experiment & & & &  &  & 1.49-2.23 \cite{Tashiro} & 1.8 \cite{Urbach} & 400 \cite{Latimer} & & & & &  \\ \hline

Ba$_{0.6}$K$_{0.4}$BiO$_3$ & 25,000 & 150 & 70 \cite{Barilo} & 3.04 \cite{BKBO} & 27.2 $\pm$ 0.2 & 4.20 $\pm$ 0.05 & 2.6 $\pm$ 0.2 & 273.3 $\pm$ 0.4 & 3.58 $\pm$ 0.05 & 2.9 & 5.3 & 10.5 & 218 \\
experiment & & & & &  & 4.5 \cite{Huang} 4.3 \cite{Kosugi} & 2.0 \cite{Woodfield} 1.8 \cite{Panova} & 289 \cite{Moseley} 270 \cite{Uchida} & & & & &  \\
  & & & & &  &  &  & 340 \cite{Ansaldo} & & & & & \\ \hline

HoNi$_{2}$B$_2$C & 20,000 & 300 & 12.5 & 1.45 \cite{YNi2B2C} & 5.7 $\pm$ 0.4 & 1.1 $\pm$ 0.2 & 2.0 $\pm$ 1.4 & 297 $\pm$ 5 & 4.4 $\pm$ 0.9 & 2.7 & 4.3 & 5.7 & 33.0 \\
experiment & & & from $\xi_0$ \cite{Wimbush} & &  & 0.95 \cite{Naidyuk} & &  & & & & & \\ \hline

YNi$_{2}$B$_2$C & 20,000 & 600 & 39 & 2.1 \cite{YNi2B2C} & 15.1 $\pm$ 0.7 & 2.6 $\pm$ 0.2 & 2.4 $\pm$ 1 & 268 $\pm$ 2 & 3.9 $\pm$ 0.3 & 7.9 & 10.6 & 9.6 & 126 \\
experiment & & & from $\xi_0$ \cite{Wimbush2} & &  & 2.2 \cite{Poole9}, 2-2.5 \cite{LuX} & 1.77 \cite{Poole9} & 350 $\pm$ 50 \cite{ProzorovLamYNi2B2C} & & & & & \\ \hline

H$_2$S (155 GPa) &  &  &  &  &  &  &  &  &  &  &  &  &  \\
4 param fit & 25,000 & 100 & 88 & 9.8 \cite{H2S} & 204.6 $\pm$ 0.1 & 26 $\pm$ 3 & 1.3 $\pm$ 0.1 & 188 $\pm$ 7 & 2.95 $\pm$ 0.3 & 6.2 & 14.9 & 23.2 & 580 \\
low-$T$ 2 param fit & & & 88 & 10.5 \cite{H2S} & & 27.8 $\pm$ 0.2 &  & 189 $\pm$ 2 & 3.17 $\pm$ 0.03 & 6.6 & 13.1 & 23.0 & 574  \\
experiment & & & from $\xi_0$ \cite{H2S} & &  &  & & 163 \cite{H2S} & & & & & \\ \hline
\end{tabular}
\caption{Fit parameters derived from $J_c(T,\textrm{sf})$ for $s$-wave superconductors (as shown in Fig.~\ref{sfits}). Measured data is shown where available. $B_x$ and $B_y$, defined in Fig.~\ref{geometry}, are the surface field components at the middle of the flat surface and edge, respectively. $B_{c1}$ and $B_c$ are calculated from $\lambda_0$. The $B_y$ and $B_{c1}$ columns are highlighted in gray for ease of comparison. }
\label{sresults}
\end{table*}

\begin{table*}[h]
\tiny
\centering
\begin{tabular}{|l||c|c|c|c|c|c|c|c|c|c|g|g|}
\hline
Material/ & $2a $ & $2b $ & $\kappa$ & $J_c(0)$ & $T_c$ & $\Delta_0$ & $\Delta C/C$ & $\lambda_0$ & 2$\Delta_0/k_B T_c$ & $B_x$ & $B_y$ & $B_{c1}$ \\
geometry & nm & nm & & MA/cm$^2$ & (K) & (meV) &  & (nm) &  & (mT) & (mT) & (mT) \\ \hline \hline
{\bf $d$-wave} & & & & &  &  &  &  &  & & & \\ \hline

HgBa$_2$CaCu$_2$O$_8$ & 450,000 & 250 & 123 \cite{Puzniak} & 10.4 \cite{Krusin} & 120 $\pm$ 1 & 16.9 $\pm$ 0.3 & 0.96 $\pm$ 0.09 & 188.3 $\pm$ 0.6 & 3.27 & 16.3 & 42.6 & 24.8 \\
experiment & & & & &  & 32 \cite{Sacuto1} &  & 145, 188 \cite{Thompson} & & & & \\ \hline

Bi$_2$Sr$_2$CaCu$_2$O$_8$ + Zn & 10,000 & 360 & 170 \cite{Ri} & 10.1 \cite{Bi2212} & 82.7 $\pm$ 2.4 & 14.0 $\pm$ 0.6 & 0.76 $\pm$ 0.17 & 196 $\pm$ 1 & 3.93 $\pm$ 0.17 & 22.8 & 29.3 & 24.2 \\
experiment & & & & &  & 20.5 \cite{Suzuki}, 23 \cite{Hudakova} & 1.5 \cite{Loram} & 180 \cite{Mosqueira} &  &  &  &  \\ \hline

Bi$_2$Sr$_2$Ca$_2$Cu$_3$O$_{10}$ & 20,000 & 100 & 170 \cite{Li} & 14 \cite{Hanisch} & 85.3 $\pm$ 0.6 & 14.2 $\pm$ 0.3 & 0.69 $\pm$ 0.04 & 175.2 $\pm$ 0.7 & 3.86 $\pm$ 0.08 & 8.8 & 16.8 & 30.2 \\
experiment & & & & &  & 30 \cite{Hudakova} &   & 151-155 \cite{Weigand2010} &  & & &  \\ \hline

Tl$_2$Ba$_2$CaCu$_{2}$O$_8$ & 12,000 & 650 & 150 \cite{Kim3} & 12.9 \cite{Tl2212} & 103 $\pm$ 1.9 & 19.1 $\pm$ 0.8 & 0.81 $\pm$ 0.13 & 179.4 $\pm$ 0.8 & 4.30 $\pm$ 0.18 & 52.7 & 61.0 & 28.2 \\
experiment & & & & &  & 16-28 \cite{Q.Huang} & 0.6 \cite{Junod} $\pm$ 0.1 & 139, 188 \cite{Moonen} & & & &  \\ \hline

(Y,Dy)Ba$_2$Cu$_3$O$_7$ & 500,000 & 850 & 95 \cite{Poole12} & 31.8 \cite{Y123ours} & 92.0 $\pm$ 2.1 & 14.9 $\pm$ 0.3 & 0.78 $\pm$ 0.14 & 122.6 $\pm$ 0.2 & 3.80 $\pm$ 0.1 & 170 & 382 & 55.4 \\
experiment & & & & &  & 16.7 \cite{Dagan} & 2.7 \cite{LoramYBCO} & 125 \cite{Sonier} & & & &  \\ \hline

(Y,Dy)Ba$_2$Cu$_3$O$_7$ & 500,000 & 1400 & 95 \cite{Poole12} & 26 \cite{YDy123} & 87.5 $\pm$ 0.4 & 12.8 $\pm$ 0.1 & 1.23 $\pm$ 0.06 & 124.3 $\pm$ 0.3 & 3.4 $\pm$ 0.05  & 229 & 478 & 53.9 \\
experiment & & & & &  & 16.7 \cite{Dagan} & 2.7 \cite{LoramYBCO} & 125 \cite{Sonier} & & & &  \\ \hline

(Nd,Eu,Gd)Ba$_2$Cu$_3$O$_7$ & 50,000 & 50 & 95 \cite{Poole12} & 30 \cite{NdEuGd123} & 86.7 $\pm$ 0.6 & 15.3 $\pm$ 0.1 & 1.8 $\pm$ 0.2 & 130.5 $\pm$ 0.2 & 4.1 $\pm$ 0.04 & 9.4 & 22.8 & 48.9 \\
experiment & & & & &  & 16.7 \cite{Dagan} & 2.7 \cite{LoramYBCO} & 118 \cite{Tallon6} & & & &  \\ \hline

GdBa$_2$Cu$_3$O$_7$ & 50,000 & 50 & 95 \cite{Poole12} & 22.7 \cite{NdEuGd123} & 85.4 $\pm$ 0.8 & 14.9 $\pm$ 0.2 & 0.75 $\pm$ 0.05 & 143.1 $\pm$ 0.2 & 4.05 $\pm$ 0.07  & 7.1 & 17.3 & 40.7 \\
experiment & & & & &  & 16.7 \cite{Dagan}] & 2.7 \cite{LoramYBCO} & 118 \cite{Tallon6} & & & &  \\ \hline

NdBa$_2$Cu$_3$O$_7$ & 5,000 & 150 & 95 \cite{Poole12} & 28.9 \cite{Nd123} & 90.9 $\pm$ 0.8 & 17.3 $\pm$ 0.4 & 1.56 $\pm$ 0.16 & 134 $\pm$ 0.4 & 4.4 $\pm$ 0.1 & 27.2 & 36.5 & 46.4 \\
experiment & & & & &  & 16.7 \cite{Dagan} & 2.7 \cite{LoramYBCO} & 118 \cite{Tallon6} & & & &  \\ \hline %\hhline{|=||=|=|=|=|=|}

\end{tabular}
\caption{Fit parameters derived from $J_c(T,\textrm{sf})$ for $d$-wave superconductors (as shown in Fig.~\ref{dfits}). Measured data is shown where available. $B_x$ and $B_y$, defined in Fig.~\ref{geometry}, are the surface field components at the middle of the flat surface and edge, respectively. $B_{c1}$ and $B_c$ are calculated from $\lambda_0$. The $B_y$ and $B_{c1}$ columns are highlighted in gray for ease of comparison. }
\label{dresults}
\end{table*}

\begin{table*}[htb]
\tiny
\centering
\begin{tabular}{|l||c|c|c|c|c|c|c|c|c|c|g|g|c|}
\hline
Material/ & $2a $ & $2b $ & $\kappa$ & $J_c(0)$ & $T_c$ & $\Delta_0$ & $\Delta C/C$ & $\lambda_0$ & 2$\Delta_0/k_B T_c$ & $B_x$ & $B_y$ & $B_{c1}$ & $B_c$ \\
geometry & nm & nm & & MA/cm$^2$ & (K) & (meV) &  & (nm) &  & (mT)  & (mT)  & (mT) & (mT) \\ \hline \hline
{\bf Two gap} & & & & & &  &  &  & & & & & \\ \hline

Al & 2,000 & 75 & 0.03 \cite{Poole12} & 6.57 \cite{Al2b} & & &  & 41.2 & & 3.1 & 3.9 & - & 4.1 \\
 First band &  &  &  & 3.57 & 1.245 $\pm$ 0.002 & 0.256 $\pm$ 0.004 & 2.58 $\pm$ 0.06 & 52.1 $\pm$ 0.3 & 4.77 $\pm$ 0.08 & & & & \\
 Second band &  &  &  & 3 & 1.137 $\pm$ 0.003 & 0.100 $\pm$ 0.001 &  3.5 $\pm$ 0.3 & 55.6 $\pm$ 0.4 & 2.04 $\pm$ 0.02 & & & & \\
experiment & & & & &  & 0.179 \cite{Wolf} & 1.45 \cite{Poole4} & 50 $\pm$ 10 \cite{Prozorov} & & & & & \\
  & & & & &  &  &  & 46-51 \cite{Doezema} 51.5 \cite{Biondi} & & & & & \\ \hline

Nb & 1,000 & 20 & 1 \cite{Kim1} & 61.2 \cite{MoGe} & & &  &  48.8 & & 7.7 & 11.3 & 34.6 & 97.8  \\
 First band &  &  &  & 44.4 & 8.28 $\pm$ 0.05 & 2.5 $\pm$ 0.3 & 1.41 $\pm$ 0.08 & 54.5 $\pm$ 0.3 & 7.1 $\pm$ 0.8 & & & & \\
 Second band &  &  &  & 16.8 & 4.26 $\pm$ 0.03 & 0.97 $\pm$ 0.06 & 3 $\pm$ 0.3 & 76.5 $\pm$ 1 & 5.3 $\pm$ 0.3 & & & & \\
experiment & & & & &  & 1.5 \cite{Poole7} & 1.93 \cite{Poole4}  & 47 \cite{Maxfield}, 59 \cite{Masuradze} & & & & & \\
experiment & & & & &  & &  & 41 $\pm$ 4 \cite{Felcher} & & & & & \\  \hline

Ba(Fe,Co)$_2$As$_2$ & 6,700 & 220 & 90 \cite{Gordon} & 2.19 \cite{BaFeCoAs} & 17.4 $\pm$ 0.2 & &  &  316 $\pm$ 1 &  & 3.0 & 4.0 & 8.3 & 210 \\
 First band &  &  &  &  & & 3.2 $\pm$ 0.4 & 1.2 $\pm$ 0.2 &  & 4.2 $\pm$ 0.6 & & & & \\
 Second band &  &  &  &  &  & 1.1 $\pm$ 0.2 &  1.43 - fixed &  & 1.5 $\pm$ 0.3 & & & & \\
 $\gamma$ = 0.76 $\pm$ 0.08 &  &  &  &  & &  &  &  &  & & & & \\
experiment & & & & &  & 3.1 \cite{Lobo} & 1.2 \cite{Gofryk} & 270 \cite{Gordon}, 274 \cite{Luan} & & & & & \\
experiment & & & & &  & &  & 307 \cite{Williams} & & & & & \\  \hline

Ba$_{0.5}$K$_{0.5}$Fe$_2$As$_2$+Zn & 119 & 102 & 90 \cite{Gordon} & 5.06 \cite{BaFeAsZn} & 21.1 $\pm$ 0.2 & &  &  294.5 $\pm$ 0.7 & & 3.2 & 1.2 & 9.5 & 242 \\
 First band &  &  &  &  & & 4.25 $\pm$ 0.03 & 2.1 $\pm$ 0.2 &  & 4.7 $\pm$ 0.3 & & & & \\
 Second band &  &  &  &  &  & 1.3 $\pm$ 0.1 &  1.43 - fixed &  & 1.5 $\pm$ 0.1 & & & & \\
 $\gamma$ = 0.74 $\pm$ 0.03 &  &  &  &  & &  &  &  &  & & & & \\
experiment & & & & &  &  & 1.65 \cite{Dong} &  & & & & & \\ \hline

FeSe$_{0.5}$Te$_{0.5}$ & 564 & 100 & 180 \cite{Bendele} & 1.52 \cite{FeSeTe} & 13.2 $\pm$ 0.2 & 3.8 $\pm$ 0.6 & 5 $\pm$ 1.4 & 451 $\pm$ 2 & 6.7 $\pm$ 1 & 0.96 & 0.76 & 4.6 & 206  \\
experiment & & & & &  & 4.1 \cite{FeSeTe}, 4.59 \cite{Perucchi} & 3 \cite{Klein} & 430 $\pm$ 50 \cite{Klein} & & & & & \\ \hline

MgB$_2$ Single-gap fit & 350 & 100 & 26 \cite{Buzea} & 84.1 \cite{Zhuang} & 39.7 $\pm$ 1.1 & 7.4 $\pm$ 0.8 & 1.0 $\pm$ 0.3 & 94.2 $\pm$ 0.3 & 4.3 $\pm$ 0.5 & 52.8 & 35 & 69.8 & 683 \\
MgB$_2$ Single-gap fit & 500 & 100 &  & 63.7 \cite{Zhuang} & 40.3 $\pm$ 0.8 & 8.1 $\pm$ 0.8 & 0.8 $\pm$ 0.1 & 100.7 $\pm$ 0.2 & 4.7 $\pm$ 0.5 & 40.0 & 30.5 & 61.0 & 597 \\

MgB$_2$ $\gamma$ = 0.64 $\pm$ 0.03 & 600 & 10 & & 128.4 \cite{MgB2a} & 35.6 $\pm$ 0.2 & & & 78.5 $\pm$ 0.2 & & 8.1 & 12.3 & 100 & 983 \\
first gap & & & & & & 6.05 $\pm$ 0.36 & 1.24 $\pm$ 0.07 & & 3.9 $\pm$ 0.2 & & & & \\
second gap & & & & & & 2.16 $\pm$ 0.15 & 1.43 - fixed & & 1.41 $\pm$ 0.09 & & & & \\

MgB$_2$ $\gamma$ = 0.81 $\pm$ 0.06 & 320 & 10 & & 121 \cite{MgB2a} & 33.9 $\pm$ 0.5 & & & 84.9 $\pm$ 0.5 &  & 7.6 & 10.1 & 85.9 & 840  \\
first gap & & & & & & 4.7 $\pm$ 0.3 & 1.4 $\pm$ 0.2 & & 3.3 $\pm$ 0.2 & & & & \\
second gap & & & & & & 1.9 $\pm$ 0.5 & 1.43 - fixed & & 1.3 $\pm$ 0.3 & & & & \\

MgB$_2$ $\gamma$ = 0.82 $\pm$ 0.03 & 5,000 & 10 & & 78.2 \cite{MgB2b} & 36.3 $\pm$ 0.4 & & & 86.7 $\pm$ 0.2 & & 4.9 & 10.8 & 82.4 & 806 \\
first gap & & & & & & 5.6 $\pm$ 0.2 & 1.5 $\pm$ 0.2 & & 3.6 $\pm$ 0.2 & & & & \\
second gap & & & & & & 1.7 $\pm$ 0.2 & 1.43 - fixed & & 1.3 $\pm$ 0.2 & & & & \\

experiment & & & & &  & 7.1/2.3 \cite{Iavarone} & 0.82-1.32 \cite{Nicol} & 85 \cite{Panagop1} & & & & & \\
& & &  & &  &  &  & 95, 100 \cite{Niedermayer} & & & & &  \\ \hline \hline %\hhline{|=||=|=|=|=|=|}
\end{tabular}
\caption{Fit parameters derived from $J_c(T,\textrm{sf})$ for two-gap superconductors (as shown in Fig.~\ref{twobfits}). Measured data is shown where available. $B_x$ and $B_y$, defined in Fig.~\ref{geometry}, are the surface field components at the middle of the flat surface and edge, respectively. $B_{c1}$ and $B_c$ are calculated from $\lambda_0$. The $B_y$ and $B_{c1}$ columns are highlighted in gray for ease of comparison. }
\label{tworesults}
\end{table*}

\begin{table*}[htb]
\tiny
\centering
\begin{tabular}{|l||c|c|c|c|c|c|c|c|c|c|g|g|c|}
\hline
Material/ & $2a $ & $2b $ & $\kappa$ & $J_c(0)$ & $T_c$ & $\Delta_0$ & $\Delta C/C$ & $\lambda_0$ & 2$\Delta_0/k_B T_c$ & $B_x$ & $B_y$ & $B_{c1}$ & $B_c$ \\
geometry & nm & nm & & MA/cm$^2$ & (K) & (meV) &  & (nm) & & (mT) & (mT) & (mT) & (mT)  \\ \hline \hline
{\bf Nanowire} & & & & &  &  &  &  & & & & &  \\ \hline

Al & 10 & 5 & 0.03 \cite{Poole12} & 9.23 \cite{Alnano} & 1.48 $\pm$ 0.01 & 0.193 $\pm$ 0.002 & 1.3 $\pm$ 0.1 & 49.33 $\pm$ 0.05 & 3.03 $\pm$ 0.05 & 0.29 & 0.15 & - & 2.87 \\
   & 9.3 & 5 &  & 8.68 \cite{Alnano} & 1.386 $\pm$ 0.007 & 0.195 $\pm$ 0.002 & 1.36 $\pm$ 0.07 & 50.4 $\pm$ 0.09 & 3.27 $\pm$ 0.05 & 0.27 & 0.14 & - & 2.75 \\
   & 8.4 & 5 &  & 7.94 \cite{Alnano} & 1.4 $\pm$ 0.01 & 0.205 $\pm$ 0.002 & 1.07 $\pm$ 0.05 & 51.89 $\pm$ 0.04 & 3.4 $\pm$ 0.06  & 0.25 & 0.12 & - & 2.60 \\
   & 7 & 5 &  & 6.4 \cite{Alnano} & 1.518 $\pm$ 0.003 & 0.182 $\pm$ 0.001 & 0.96 $\pm$ 0.01 & 55.769 $\pm$ 0.007 & 2.78 $\pm$ 0.02  & 0.20 & 0.09 & - & 2.25 \\
   & 5.4 & 5 &  & 4.94 \cite{Alnano} & 1.13 $\pm$ 0.01 & 0.14 $\pm$ 0.001 & 2.3 $\pm$ 0.3 & 60.79 $\pm$ 0.06 & 2.88 $\pm$ 0.05  & 0.16 & 0.06 & - & 1.89  \\
experiment & & & & &  & 0.179 \cite{Wolf} & 1.45 \cite{Poole4} & 50 $\pm$ 10 \cite{Prozorov} & & & & & \\
  & & & & &  &  &  & 46-51 \cite{Doezema} 51.5 \cite{Biondi} & & & & & \\ \hline

Zn & 100 & 65 & 3 & 0.0302 \cite{Znnano} & 0.82 $\pm$ 0.03 & 0.23 $\pm$ 0.04 & 4.3 $\pm$ 1.7 & 1114 $\pm$ 5 & 6.5 $\pm$ 1.1 & 0.012 & 0.006 & 0.21 & 0.56 \\
Zn & 100 & 65 &   & 0.0800 \cite{ZnnanoPRB} & 0.836 $\pm$ 0.02 & 0.228 $\pm$ 0.006 & 4.6 $\pm$ 0.7 & 805.9 $\pm$ 0.3 & 6.3 $\pm$ 0.2 & 0.033 & 0.015 & 0.41 & 1.08 \\
experiment & & & from $H_{c0}$ \cite{Rothwarf} & &  & 0.115 \cite{Poole7} &  & 30-60.5 \cite{Bonaldea} & & & & &  \\ \hline

Amorph-W & 250 & 50 & 117 & 0.34 \cite{AmorphW} & 4.7 $\pm$ 0.2 & 0.9 $\pm$ 0.2 & 1.2 $\pm$ 0.5 & 738 $\pm$ 13 & 4.6 $\pm$ 1.1 & 0.11 & 0.08 & 1.59 & 50.0 \\
experiment & & & from $\xi_0$ \cite{AmorphW} & & & 0.66 \cite{Guillamon}  &  &  & & & & & \\ \hline

PbIn (cylindrical) & 27.5 & - & 5 \cite{Poole12} & 1.06 \cite{PbInnano} & 7.6 $\pm$ 0.6 & 1.6 $\pm$ 0.1 & 5 $\pm$ 4 & 295 $\pm$ 2 & 4.8 $\pm$ 0.4 & 0.09 & 0.09 & 4.0 & 13.4 \\
experiment & & & & &  & 1.2-1.25 \cite{Rao} & 2.5 \cite{Padamsee} & 150 \cite{Poole12} & & & & &  \\ \hline

MoN (cylindrical) & 320 & - & 54 & 0.58 \cite{MoNnano} & 12.6 $\pm$ 0.1 & 2.9 $\pm$ 0.1 & 5 $\pm$ 0.7 & 463 $\pm$ 1 & 5.3 $\pm$ 0.2  & 0.58 & 0.58 & 3.45 & 58.7 \\
experiment & & & from $\xi_0$ \cite{Christen} & &  & 1.94 \cite{Groll} &  & 440 $\pm$ 40 \cite{MoN} & & & & & \\ \hline

YBa$_2$Cu$_3$O$_7$ & 136 & 50 & 95 \cite{Poole12} & 17.5 \cite{Larsson} & 81 $\pm$ 1 & 15.3 $\pm$ 0.3 & 1.4 $\pm$ 0.2 & 160.7 $\pm$ 0.3 & 4.38 $\pm$ 0.01 & 5.5 & 3.3 & 32.2 & 857 \\

experiment & & & & &  & 16.7 \cite{Dagan} & 2.7 \cite{LoramYBCO} & 125 \cite{Sonier} & & & & & \\ \hline

\end{tabular}
\caption{Fit parameters derived from $J_c(T,\textrm{sf})$ for nanowires (as shown in Fig.~\ref{nanowires}). Measured data is shown where available. $B_x$ and $B_y$, defined in Fig.~\ref{geometry}, are the surface field components at the middle of the flat surface and edge, respectively. $B_{c1}$ and $B_c$ are calculated from $\lambda_0$. The $B_y$ and $B_{c1}$ columns are highlighted in gray for ease of comparison. }
\label{nanoresults}
\end{table*}

%\section*{References}

\end{document}